\documentclass[nobibnotes,superscriptaddress,nofootinbib,twocolumn]{revtex4-1}

\usepackage{amsmath,amssymb,bbm}
\usepackage{graphicx}

\DeclareMathOperator{\diag}{diag}
\DeclareMathOperator{\imag}{Im}

\begin{document}

\title{Leptonic CP Violation from a New Perspective}
\author{David Emmanuel-Costa}
\email{david.costa@tecnico.ulisboa.pt}
\affiliation{Departamento de F\'{\i}sica and CFTP, Instituto Superior T\'{e}cnico (IST),
 Universidade de Lisboa, Av. Rovisco Pais 1, 1049-001 Lisboa, Portugal}

\author{Nuno Rosa Agostinho}
\email{nuno.agostinho@tecnico.ulisboa.pt}
\affiliation{Departamento de F\'{\i}sica and CFTP, Instituto Superior T\'{e}cnico (IST),
 Universidade de Lisboa, Av. Rovisco Pais 1, 1049-001 Lisboa, Portugal}

\author{J.I. Silva-Marcos}
\email{juca@cftp.tecnico.ulisboa.pt}
\affiliation{Departamento de F\'{\i}sica and CFTP, Instituto Superior T\'{e}cnico (IST),
 Universidade de Lisboa, Av. Rovisco Pais 1, 1049-001 Lisboa, Portugal}

\author{Daniel Wegman}
\email{wegman@cftp.tecnico.ulisboa.pt}
\affiliation{Departamento de F\'{\i}sica and CFTP, Instituto Superior T\'{e}cnico (IST),
 Universidade de Lisboa, Av. Rovisco Pais 1, 1049-001 Lisboa, Portugal}


\preprint{CFTP/15-005}

\begin{abstract}
We study leptonic CP violation from a new perspective. For Majorana
neutrinos, a new parametrization for leptonic mixing of the form $V=O_{23}\,O_{12}\, 
K_{a}^{i}\cdot O$ reveals interesting aspects that are
less clear in the standard parametrization. We identify several important
scenario-cases with mixing angles in agreement with experiment and leading
to large leptonic CP violation. If neutrinos happen to be quasi-degenerate,
this new parametrization might be very useful, e.g., in reducing the number
of relevant parameters of models.
\end{abstract}

\maketitle


\section{Introduction}

Observations of neutrino oscillations have solidly established the
massiveness of the neutrinos and the existence of leptonic mixing. Since
neutrinos are strictly massless in the standard model (SM), these
observations require necessarily new physics beyond the SM. One is still far
from a complete picture of the lepton sector, i.e., many fundamental
questions need to be answered. Not only the origin of the leptonic flavor
structure remains unknown, but leptonic mixing differs tremendously from the
observed quark mixing. Moreover, the absolute neutrino mass scale is still
missing, one does not know whether neutrinos are Majorana or Dirac
particles, and the nature of leptonic CP violation is still open (for a
recent review see Ref.~\cite{Branco:2011zb}).

During the last decades, several attempts were made in order to overcome
these fundamental questions. In particular, one may impose family symmetries
forbidding certain couplings and at the same time explaining successfully
the observed structure of masses and mixings, as well as predicting some
other observables~\cite{King:2014nza,Shimizu:2014ria,Wang:2013wya,King:2013eh,Boucenna:2012xb,Ferreira:2012ri,Branco:2012vs,Hernandez:2012sk,Eby:2011aa,Ge:2011qn,Toorop:2011jn,Zhou:2011nu,Altarelli:2010gt,Datta:2005ci}. Although the structure of leptonic mixing is predicted in such models, the
mass spectrum turns out to be unconstrained by such symmetries. The
connection of leptonic mixing angles and CP-phases with neutrino spectra in
the context of partially and completely degenerate neutrinos was proposed in~\cite{Hernandez:2013vya}. In an alternative approach, the anarchy of the
leptonic parameters is assumed so that there is no physical distinction
among three generations of lepton doublets~\cite{Hall:1999sn,Haba:2000be,deGouvea:2003xe,Altarelli:2012ia}.

From the analysis of neutrino oscillation experiments one can extract bounds
for the light neutrino mass square differences $\Delta m_{21}^{2}\equiv
m_{2}^{2}-m_{1}^{2}$ and $\Delta m_{31}^{2}\equiv m_{3}^{2}-m_{1}^{2}$. All
knowledge on the light neutrino mixing is encoded in the
Pontecorvo-Maki-Nakagawa-Sakata matrix (PMNS)~\cite{Pontecorvo:1957cp,Maki:1962mu,Pontecorvo:1967fh}. In order to further
analyze the leptonic flavour structure, it is essential to parametrize all
the entries of the full PMNS matrix in terms of six independent parameters.
It is clear that the choice of a parametrization does not impose any
constraints on the physical observables. However, parametrizations are an important
tool to interpret underlying symmetries or relations that the data may
suggest. In this sense, different parametrizations are certainly equivalents
among themselves, although some particular patterns indicated by the data
are easer to visualize in some parametrizations than in others. Moreover,
special limits suggested by some parametrizations are obfuscated in others.

Among many parametrization proposed in the literature, the standard
parametrization is the most widely used, and the six parameters are three
mixing angles, namely $\theta_{12},\,\theta_{13}\,,\theta_{23}\in \lbrack
0,\pi /2]$, one Dirac-type phase $\delta$ and two Majorana phases $\alpha
_{1},\alpha_{2}$ in following form: 
\begin{equation}
V^{SP}\,=\,K\cdot O_{23}\cdot K_{D}\cdot O_{13}\cdot O_{12}\cdot K_{M}\,,
\label{eq:spParm}
\end{equation}
where the real orthogonal matrices $O_{12}$, $O_{13}$ and $O_{23}$ are the
usual rotational matrices in the $(1,2)-$, $(1,3)-$, and $(2,3)-$sector,
respectively. The diagonal unitary matrices $K_{D}$ and $K_{M}$ are given by 
$K_{D}\equiv \diag(1,1,e^{i\alpha_{D}})$ and $K_{M}\equiv \diag(1,e^{i\alpha_{1}^{M}},
e^{i\alpha_{2}^{M}})$. Within the standard
parametrization, one may recall that the consistent values for the neutrino
mixing angles $\theta_{12}$ and $\theta_{23}$ together with the smallness
of $\theta_{13}$ suggest that the neutrino mixing is rather close to the
tribimaximal mixing (TBM)~\cite{Harrison:2002er}. It is important to stress
that this parametrization is (modulo irrelevant phases), the same as the one
used for the quark sector, despite the fact of leptonic mixing being quite
different.

In this paper we study leptonic CP Violation in the context of a new
parametrization for leptonic mixing of the form 
\begin{equation}
V\,=\,O_{23}\,O_{12}\,K_{\alpha}^{i}\cdot O\,,  \label{newparam}
\end{equation}
where $O$ is a real orthogonal matrix parametrized with three mixing angles.
This new parametrization turns out to be very useful in the case where
neutrinos are quasi-degenerate Majorana fermions~\cite{Branco:1998bw}. It may
reflect some specific nature of neutrinos, suggesting that there is some
major intrinsic Majorana character of neutrino mixing and CP violation
present in the left part of the parametrization, while the right part, in
the form of the orthogonal matrix $O$, may reflect the fact that there are 3
neutrino families with small mass differences and results in small mixings.
Thus, the intrinsic Majorana character of neutrinos may be large with a
large contribution to neutrino mixing (from some yet unknown source), while
the extra mixing $O$ of the families is comparable to the quark sector and
may be small, of the order of the Cabibbo angle.

The new parametrization permits a new view of large leptonic CP Violation.
It reveals interesting aspects that are less clear in the standard
parametrization. We identify five scenario-cases that lead to large Dirac-CP
violation, and which have mixing angles in agreement with experimental data.
A certain scenario (I-A) is found to be the most appealing, since it only
needs 2 parameters to fit the experimental results on lepton mixing and
provides large Dirac-CP violation and large values for the Majorana-CP
violating phases.

The paper is organized as follows. In the next section, we prove the
consistency of the new parametrization stated in Eq.~\eqref{newparam}. In
Sec.~\ref{deg1}, we motivate the use of this new parametrization in the
limit of degenerate or quasi-degenerate neutrino spectrum. Then in 
Sec.~\ref{lcpv}, we present an alternative view of large leptonic CP violation, using
the new parametrization for leptonic mixing, discuss its usefulness and
identify several important scenario-cases. Results are shown for mixing and
CP violation. In Sec. \ref{num}, we give a numerical analysis of the
scenarios described in the previous section, and, for the quasi-degenerate
Majorana neutrinos, a numerical analysis of their stability. Finally, in
Sec.~\ref{sec:con}, we present our conclusions.

\section{A novel parametrization\label{np}}

In this section, we present the new parametrization for the lepton mixing
matrix. First, we prove that any unitary matrix can be written with the
following structure: 
\begin{equation}
V\,=K_{S}\,\ O_{23}\,O_{12}\ K_{\alpha}^{i}\cdot O\,,  \label{new}
\end{equation}
where $K_{S}\,=\diag(e^{i\alpha_{1}},e^{i\alpha_{2}},e^{i\alpha_{3}})$ is
a pure phase unitary diagonal matrix, $O_{23},$ $O_{12}$ are two elementary
orthogonal rotations in the $(23)$- and $(12)$-planes, $K_{\alpha}^{i}=
\diag(1,i,e^{i\alpha})$ has just one complex phase $\alpha$ (apart from the
imaginary unit $i$), and $O$ is a general orthogonal real matrix described
by 3 angles.

Proof: Let us start from a general unitary matrix $V$ and compute the
following symmetric unitary matrix $S$, 
\begin{equation}
S\,=\,V^{\ast}\,V^{\dagger}\,.  \label{s}
\end{equation}
Assuming that $S\,$\ is not trivial, i.e. it is not a diagonal unitary
matrix, one can rewrite the matrix $S$ as 
\begin{equation}
S\,=\,K_{S}^{\ast}\,\,\ S_{0}\,\ K_{S}^{\ast}\,,  \label{s1}
\end{equation}
with a pure phase diagonal unitarymatrix $K_{S}=\diag(e^{i\alpha
_{1}},e^{i\alpha_{2}},e^{i\alpha_{3}})$ so that the first row and the
first column of $S_{0}$ become real. In fact, the diagonal matrix $K_{S}$
has no physical meaning, since it only rephases the PMNS matrix $V$ on the
left. This can be clearly seen in the weak basis where the charged lepton
mass matrix is diagonal and through a weak basis transformation the phases
in $K_{S}$ can be absorbed by the redefinition of the right-handed charged
lepton fields. One can now perform a $(23)-$rotation on $S_{0}$ as, 
\begin{equation}
S_{0}^{\prime}\,=\,O_{23}^{\intercal}\,S_{0}\,O_{23}\,,
\end{equation}
with a orthogonal matrix $O_{23}$ given by 
\begin{equation}
O_{23}\,=\,
\begin{pmatrix}
1 & 0 & 0 \\ 
0 & \cos \theta_{23} & \sin \theta_{23} \\ 
0 & -\sin \theta_{23} & \cos \theta_{23}
\end{pmatrix}
\,,
\end{equation}
so that the $(13)-$ and $(31)-$elements of the resulting matrix vanish.
Making use of unitarity conditions one concludes that automatically the 
$(13)-$ and $(31)-$elements become also zero and therefore the $(12)-$sector
of $S_{0}^{\prime}$ decouples and one obtains that 
\begin{equation}
S_{0}\,=\,O_{23}\,O_{12}\cdot \diag(1,-1,e^{-2i\alpha})\cdot
O_{12}^{\intercal}\,O_{23}^{\intercal}\,,
\end{equation}
where $O_{12}$ is given by 
\begin{equation}
O_{12}\,=\,
\begin{pmatrix}
\cos \theta_{12} & \sin \theta_{12} & 0 \\ 
-\sin \theta_{12} & \cos \theta_{12} & 0 \\ 
0 & 0 & 1
\end{pmatrix}
\,.
\end{equation}
The matrix $S_{0}$ is then written as 
\begin{equation}
S_{0}\,=\,U_{0}^{\ast}\,U_{0}^{\dagger}\,,  \label{s0}
\end{equation}
where the unitary matrix $U_{0}$ is given by 
\begin{equation}
U_{0}\,=\,O_{23}\,O_{12}\,K_{\alpha}^{i}\,,  \label{uo}
\end{equation}
with $K_{\alpha}^{i}=\diag(1,i,e^{i\alpha})$. Thus, given $V$, we can
compute explicitly the matrices $K_{S}$, $O_{23}$, $O_{12}$ and $K_{\alpha
}^{i}$. In order to obtain the form given in Eq.~\eqref{new}, we factorize
the leptonic mixing $V$ as 
\begin{equation}
V\,=\,(K_{S}\,U_{0})\,W_{0}^{\ast}\,,
\end{equation}
and we demonstrate that $W_{0}$ is real and orthogonal. By definition, the
matrix $W_{0}$, 
\begin{equation}
W_{0}\,\equiv \,\left( U_{0}^{\intercal}\ K_{S}\right) \cdot V^{\ast}\,,
\label{W}
\end{equation}
is obviously unitary since it is the product of unitary matrices. Let us
then verify that $W_{0}$ is indeed orthogonal by computing the product 
\begin{equation}
\begin{split}
W_{0}\cdot W_{0}^{\intercal}& =\left( U_{0}^{\intercal}\,K_{S}\right)
\cdot V^{\ast}\ V^{\dagger}\cdot \left( K_{S}\,U_{0}\right) \\
& =U_{0}^{\intercal}\,K_{S}\,\ S\,\ K_{S}\ U_{0}\,,
\end{split}
\end{equation}
where we have used Eq.~\eqref{s}. Inserting into this expression the other
expression for $S$ given in Eq.~\eqref{s1},and making use of Eq.~\eqref{s0} we
find 
\begin{equation}
W_{0}\cdot W_{0}^{\intercal}=\mathbbm{1}\,,  \label{W2}
\end{equation}
which means that $W_{0}$ is real and orthogonal. We thus write $W_{0}$
explicitly as 
\begin{equation}
W_{0}\equiv \left( U_{0}^{\intercal}\,K_{S}\right) \cdot V^{\ast}\equiv
O\,,  \label{o2}
\end{equation}
where the $O$ stands for the fact that it is an real-orthogonal matrix.
Finally, rewriting this equation, we find for the general unitary matrix $V$ 
\begin{equation}
V\,=\,K_{S}\cdot U_{0}\cdot O\,,  \label{v3}
\end{equation}
or with Eq.~\eqref{uo} 
\begin{equation}
V\,=\,K_{S}\cdot O_{23}\,O_{12}\,K_{\alpha}^{i}\cdot O\,.  \label{v4}
\end{equation}
We have thus derived a new parametrization for the lepton mixing matrix,
i.e., 
\begin{equation}
V\,=\,O_{23}\,O_{12}\,K_{\alpha}^{i}\cdot O\,,  \label{pmns}
\end{equation}
where we have discarded the unphysical pure phase matrix $K_{S}$. It is
clear that, as with the standard parametrization in Eq.~\eqref{eq:spParm},
this parametrization has also 6 physical parameters, but, some are now of a
different nature: 2 angles in $O_{23}$ and $O_{12}$, 3 other angles in $O$,
but just one complex phase $\alpha $ in $K_{\alpha}^{i}$. From now on, we
use explicitly the following full notation 
\begin{equation}
V\,=\,O_{23}^{L}\,O_{12}^{L}\,\cdot K_{\alpha}^{i}\cdot
O_{23}^{R}\,O_{13}^{R}\,O_{12}^{R}\,,  \label{pmns1}
\end{equation}
where we have identified each of the elementary orthogonal rotations, either
on the left or on the right of the CP-violating pure phase matrix $K_{\alpha
}^{i}$, with a notation superscript $L,R$.

\subsubsection{Other parametrizations}

For completeness, we point out, following a similar procedure outlined here,
that one can also obtain other forms (from the one in Eqs.~\eqref{pmns} and~\eqref{pmns1}) 
for the parametrization of the lepton mixing matrix\cite{SilvaMarcos:2002nn}. E.g. one can 
have a parametrization where $V=O_{23}\,O_{13}\,K_{i}^{\alpha}\cdot O$, with $K_{i}^{\alpha}=\diag
(1,e^{i\alpha},i)$, or even other variations such as $V=O_{12}\,O_{23}\,
\widetilde{K}_{i}^{\alpha}\cdot O$, with $\widetilde{K}_{i}^{\alpha}=\diag(e^{i\alpha},i,1)$. 
Here, we concentrate on the parametrization of Eqs. (\ref{pmns}, \ref{pmns1}) and discuss its particular usefulness.

\subsubsection{General formulae}

The angles $\theta_{23}^{L}$ and $\theta_{12}^{L}$ can be easily
calculated from the PMNS matrix $V$ as 
\begin{equation}
\left\vert \tan \theta_{23}^{L}\right\vert \,=\,\frac{r_{3}}{r_{2}}\,,\quad
\left\vert \tan 2\theta_{12}^{L}\right\vert \,=\,\frac{(r_{2}^{2}+r_{3}^{2})\left\vert \cos \theta_{23}^{L}\right\vert}{r_{1}r_{2}}\,,
\end{equation}
where the real numbers $r_{1}$, $r_{2}$ and $r_{3}$ are given by 
\begin{align}
r_{1}& =\left\vert V_{11}^{2}\,+\,V_{12}^{2}\,+\,V_{13}^{2}\right\vert \,, \\
r_{2}& =\left\vert V_{11}V_{21}\,+\,V_{12}V_{22}\,+\,V_{13}V_{23}\right\vert
\,, \\
r_{3}& =\left\vert V_{11}V_{31}\,+\,V_{12}V_{32}\,+\,V_{13}V_{33}\right\vert
\,.
\end{align}
The phase $\alpha $ in $K_{i}^{\alpha}$ is given by $\arg
[(O_{23}^{\intercal}\,S_{0}\,O_{23})_{33}]$.

\subsubsection{CP violation}

It is worth to note that even when $\alpha=0$ or $\pi$, we still have CP
violation due to the presence of an imaginary unit in the diagonal matrix 
$K_{i}^{\alpha}$. In particular, setting $\alpha=0$ the Dirac CP violation
invariant $I_{CP}\equiv\imag(V_{12}V_{23}V_{22}^{\ast}V_{13}^{\ast})$
yields: 
\begin{widetext}
\begin{equation}
\begin{split}
I_{CP}&=\frac{1}{32} (\sin2\theta^L_{23}\cos2\theta^R_{23} (\sin^2\theta^L_{12} \cos\theta^L_{12} 
\sin2\theta^R_{12} (3 \sin3\theta^R_{13}-5 \sin\theta^R_{13})\\
&+ 8\sin^2\theta^L_{12}\cos\theta^L_{12} \cos2\theta^R_{12}\cos2\theta^R_{13}\sin2\theta^R_{23}
+(7 \cos\theta^L_{12}+\cos3\theta^L_{12}) \sin2\theta^R_{12} 
\sin\theta^R_{13} \cos^2\theta^R_{13})\\
&+2\sin2\theta^L_{12} \cos2\theta^R_{23} 
\cos\theta^R_{23} \left(\sin2\theta^R_{12} \cos\theta^R_{13} (\cos2\theta^R_{13}-3)
  \cos2\theta^R_{23}+2\cos^2\theta^R_{13}\right)
  -2 \cos2\theta^R_{12} 
  \sin2\theta^R_{13} \sin2\theta^R_{23}))\,,
\end{split}
\end{equation}
\end{widetext}
which vanishes when $\theta^L_{12}=\theta^L_{23}=0$ (i.e, omitting the left
orthogonal matrices in Eq.~\eqref{pmns1}) and when $\theta^L_{12}=
\theta^R_{23}=0$.

\subsubsection{Usefulness}

Why a new parametrization? Does it add anything useful to the standard
parametrization? We give several motivations.

First, we still do not know whether neutrinos are hierarchical, or
quasi-degenerate. However, if neutrinos happen to be quasi-degenerate, then
the new parametrization is very useful.

Secondly and in this case, the new parametrization may reflect some
intrinsic nature of neutrinos. Heuristically, it may suggest that there is
some major intrinsic Majorana character of neutrino mixing and CP violation,
present in the left part $O_{23}^{L}\,O_{12}^{L}\,K_{\alpha}^{i}$ of Eq.~\eqref{pmns1},
 while the right part in the form of the real-orthogonal
matrix $O=O_{23}^{R}\,O_{13}^{R}\,O_{12}^{R}$ with the 3 angles, may reflect
the fact that there are 3 neutrino families with small mass differences and
results in small mixing. Thus, the intrinsic Majorana character of neutrinos
may be large with large contribution to neutrino mixing (maybe from some yet
unknown source), while the extra mixing $O$ of the families is comparable to
the quark sector and may be small, of the order of the Cabibbo angle.

The third motivation is that this parametrization permits a different view
of large leptonic CP violation from a new perspective. It reveals
interesting aspects that were less clear in the standard parametrization.
The Dirac and Majorana CP violation quantities are here simply related to
just one complex phase $\alpha$ present in $K_{\alpha}^{i}=\diag(1,i,e^{i\alpha})$. 
We discuss these issues in the next subsection, first
in the limit of degenerate and quasi-degenerate Majorana Neutrinos.


\section{Degenerate and Quasi-degenerate Majorana Neutrinos\label{deg1}}

\subsection{Degenerate neutrino masses}

In the weak basis where the charged lepton mass matrix is diagonal and
real-positive, the matrix $S_{0}$ has a special meaning in the limit of
exact neutrino mass degeneracy~\cite{Branco:1998bw,Branco:2014zza}. In this
limit the neutrino mass matrix $M_{0}$ assumes the following form: 
\begin{equation}
M_{0}\,=\,{\mu}\,S_{0}\,=\,{\mu}\,U_{0}^{\ast}\,U_{0}^{\dagger}\,,
\label{mat}
\end{equation}
where $\mu$ is the common neutrino mass. The matrix $U_{0}$ accounts for
the leptonic mixing. Thus, within the parametrization given in Eq.~\eqref{pmns}, 
degeneracy of Majorana neutrino masses corresponds to setting
the orthogonal matrix $O$ to the identity matrix. In the limit of exact
degenerate neutrinos, the orthogonal matrix $O$ on the right of the new
parametrization in Eq.~\eqref{pmns}, has no physical meaning. It can be
absorbed in the degenerate neutrino fields. This has motivated our proposal
for the use of the new parametrization.

As stated in Ref.~\cite{Branco:1998bw}, in the limit of exact degeneracy for
Majorana neutrinos, leptonic mixing and CP violation can exist irrespective
of the nature of neutrinos. Leptonic mixing can only be rotated away, if and
only if, there is CP invariance and all neutrinos have the same CP parity~\cite{Wolfenstein:1981rk},
 \cite{Branco:1999fs}. This is clearly the case
when $S_{0}$ is trivial. It is also clear that even in the limit of exact
degeneracy with CP conservation, but with different CP-parities ($\alpha =0$
or $\alpha =\frac{\pi}{2}$), one cannot rotate $U_{0}$ away through a
redefinition of the neutrino fields. Thus even in this limit (within the
degeneracy limit), leptonic mixing may occur.

\subsection{Quasi-degenerate neutrinos masses}

The usefulness of the new parametrization is particulary interesting if
neutrinos are quasi-degenerate. When the degeneracy is lifted, i.e. for
quasi-degenerate neutrinos, the full neutrino mass matrix becomes slightly
different from the exact limit in Eq.~\eqref{mat}: 
\begin{equation}
M\,=\,{\mu}\ \left( S_{0}+Q^{\varepsilon}\right) \,,  \label{mat1}
\end{equation}
where $Q^{\varepsilon}$ is some small perturbation. In general, this
perturbation may significantly modify the mixing result for the exact case
in Eq.~\eqref{mat}. In view of our new parametrization, now the full lepton
mixing matrix diagonalizing $M$ is described by 
\begin{equation}
V\,=\,U_{o}^{\prime}\cdot O\,,  \label{pmns2}
\end{equation}
where $U_{o}^{\prime}$ is of the same form as $U_{o}$. It is not guaranteed
that this $U_{o}^{\prime}$ is the exactly same as $U_{o}$. It may differ
from $U_{o}$ because of the perturbation, just as the matrix $O$, which can
either small or possibly some large general orthogonal matrix. In Sec.~\ref{num},
we shall quantify this more explicitly, using numerical simulations.

\subsection{CP Violation of Quasi-degenerate Neutrinos}

It was pointed out in Ref.~\cite{Branco:1998bw}, that if neutrinos are
quasi-degenerate (or even exact degenerate) CP violation continues to be
relevant. This can be understood if one defines Weak-Basis invariant
quantities sensitive to CP violation. An important invariant quantity, in
this case, is 
\begin{equation}
G_{m}\equiv \left\vert Tr\left( \left[ M_{v}\ H_{l}\ M_{v}^{\ast
},\,H_{l}^{\ast}\right] ^{3}\right) \right\vert \,,  \label{g}
\end{equation}
where $H_{l}=M_{l}M_{l}^{\dagger}$ is the squared charged lepton mass
matrix. Contrary to the usual quantity $I=Tr\left( [M_{v}^{\dagger
}M_{v},H_{l}]^{3}\right)$ which is proportional to the Dirac CP violation
quantity $I_{CP}$, we find that the quantity $G_{m}$ signals CP violation
even if neutrinos are exact degenerate. In fact, we obtain in this limit 
\begin{equation}
G\equiv \frac{G_{m}}{\Delta_{m}}=\frac{3}{4}\,\left\vert \sin 2\theta
_{12}^{L}\sin 4\theta_{12}^{L}\sin ^{2}2\theta_{23}^{L}\sin 2\alpha
\right\vert \,,  \label{g3}
\end{equation}
where 
\begin{equation}
\Delta_{m}\,\equiv \,\mu ^{6}(m_{\tau}^{2}-m_{\mu}^{2})^{2}(m_{\tau
}^{2}-m_{e}^{2})^{2}(m_{\mu}^{2}-m_{e}^{2})^{2}\,,
\end{equation}
with $\mu$ the common neutrino mass. $\theta_{12}^{L}$ and $\theta
_{23}^{L}$ are, respectively, the angles of $O_{12}^{L}$ and $O_{23}^{L}$ in
Eq.~\eqref{pmns1}), and $\alpha$ is the complex phase of $K_{\alpha}^{i}=
\diag(1,i,e^{i\alpha})$. Obviously, with the new parametrization for the
lepton mixing in Eq.~\eqref{pmns1}, this invariant takes on a new and
relevant meaning. It is a curious fact that $G$ is so specifically (and in
such a clean way) dependent on only, what we have called, the left part of
Eq.~\eqref{pmns1} and on $\sin 2\alpha$. One is tempted to wonder whether
there could be processes directly related to this combined CP violation
quantity, instead of the usual Dirac or Majorana effects of CP violation.

\subsection{Quasi-degenerate Neutrinos and Double Beta-Decay}

Another result which we obtain in the case of quasi-degenerate neutrinos, is
the fact that the parameter $M_{ee}$ measuring double beta-decay, depends in
our new parametrization mainly on the matrix $U_{o}$. From Eq.~\eqref{mat1},
it is clear that 
\begin{equation}
\left\vert M_{ee}\right\vert \,=\,\left\vert {\mu}\ \left( S_{o}\right)
_{11}\right\vert =\left\vert {\mu}\ \cos 2\theta_{12}^{L}\right\vert \,,
\label{s11}
\end{equation}
in zeroth order in $\varepsilon$. This is an interesting result for $M_{ee}$
when confronting it with the one calculated directly from the standard
parametrization in Eq.~\eqref{eq:spParm}. In the case of quasi-degenerate
neutrinos, we have the approximation 
\begin{equation}
\left\vert M_{ee}\right\vert \,=\,\left\vert {\mu}\,\left( \cos ^{2}\theta
_{sol}+e^{2i\alpha_{1}^{M}}\sin ^{2}\theta_{sol}\right) \right\vert \,,
\label{mee1}
\end{equation}
neglecting the terms with $V_{13}^{2}$.

The point here is that, with possible separate future results for $\mu$ and 
$M_{ee}$, we may deduce if there is any significant Majorana-type phase $\alpha_{1}^{M}$.
Subsequently, by comparing Eq.~\eqref{mee1} with Eq.\eqref{s11}, we may know if 
$\theta_{12}^{L}$ can be identified with the
solar mixing angle $\theta_{sol}$. If however, this is not the case, then
we also know that the perturbation in Eq.~\eqref{mat} produces large
effects. E.g. suppose that inserting the (future) experimental results in
Eq.~\eqref{mee1} yields $\alpha_{1}^{M}=0$, which from Eq.~\eqref{s11}
results in $\theta_{12}^{L}=0$. Then a large solar angle must come mainly
from the $O$ in Eq.~\eqref{pmns}.


\section{Leptonic CP violation from a New Perspective \label{lcpv}}

Maximum Dirac-CP violation in lepton mixing can be obtained in the Standard
Parametrization of Eq.~\eqref{eq:spParm} when choosing the (Dirac) phase 
$\alpha_{D}=\pi /2$ in the diagonal unitary matrix $K_{D}=\diag(1,1,e^{i\alpha_{D}})$.
 If neutrinos are Dirac, then there is no other form
of leptonic CP violation. If neutrinos are Majorana, then there are 2 more
CP violation phases in $K_{M}=\diag(1,e^{i\alpha_{1}^{M}},e^{i\alpha
_{2}^{M}})$. These Majorana phases may be large or small, and one finds that
leptonic CP violation is apparently limited to these two considerations if
one chooses the Standard Parametrization. On the contrary, if one switches
to the new parametrization of Eq.~\eqref{pmns1}, one gets a much richer
structure for leptonic CP violation, particularly, if neutrinos are
quasi-degenerate.

The experimentally measured mixing angles are given by the paramenters of
the new parametrization as: 
\begin{widetext}
\begin{subequations}
\begin{align}
\left\vert V_{13}\right\vert^{2} &= s^2_{\theta^L_{12}} c^2_{\theta^R_{13}} s^2_{\theta^R_{23}} 
\,+\, c^2_{\theta^L_{12}} s^2_{\theta^R_{13}}\,,\\[2mm] 
\sin^2\theta_{sol} & = \frac{s^2_{\theta^L_{12}} \left(c_{\theta^R_{12}} c_{\theta^R_{23}}-s_{\theta^R_{12}} s_{\theta^R_{13}} 
s_{\theta^R_{23}}\right)^2+c^2_{\theta^L_{12}} s^2_{\theta^R_{12}} c^2_{\theta^R_{13}}}{1\,-\,\left\vert V_{13}\right\vert^{2}}\,,\\[2mm] 
\sin^2\theta_{atm} & = \frac{c^2_{\alpha} c^2_{\theta^R_{13}} c^2_{\theta^R_{23}} s^2_{\theta^L_{23}} 
 - 2 c_{\alpha} c_{\theta^L_{23}} c_{\theta^R_{13}} c_{\theta^R_{23}} s_{\theta^L_{12}} s_{\theta^L_{23}} s_{\theta^R_{13}} + 
 c^2_{\theta^L_{23}} s^2_{\theta^L_{12}} s^2_{\theta^R_{13}} + 
 c^2_{\theta^R_{13}} \left(s_{\alpha} c_{\theta^R_{23}} s_{\theta^L_{23}} + c_{\theta^L_{12}} c_{\theta^L_{23}} s_{\theta^R_{23}}\right)^2
}{1\,-\,\left\vert V_{13}\right\vert^{2}}\,.
\end{align}
\end{subequations}
\end{widetext}where we have used the identification $c_{X}=\cos X$ and 
$s_{X}=\sin X$. As will be shown, these expressions simplify significantly
for several cases near to experimental data and with large leptonic CP
violation.

Next, we identify these important cases leading to large CP violation in
lepton mixing using the new parametrization of Eq.~\eqref{pmns1}. We do this
by fixing some of the parameters, and assume this fixing would arise from a
preexisting model and/or symmetry. We choose a starting point for the mixing
matrix that has the same mixing angles as the tribimaximal mixing, 
\begin{equation}
|V_{13}|^{2}=0,\;\;sin^{2}\theta_{atm}=1/2,\;\;sin^{2}\theta_{sol}=1/3.
\label{iTBM}
\end{equation}
These values are close to the experimental results at one-sigma 
level~\cite{Forero:2014bxa}, 
\begin{equation}
\begin{aligned} 0.439 &<\sin^2 \theta_{23}< 0.599\,, \\ 0.0214 &<\sin^2
\theta_{13}< 0.0254\,, \\ 0.307 &<\sin^2 \theta_{12l}< 0.339\,. \end{aligned}
\label{tortola}
\end{equation}
given in terms of the Standard Paramatrization angles. Is easy to observe
that 1/3 is an allowed value for $\sin ^{2}\theta_{12}$, but values
slightly lower are better. The central value for $\sin ^{2}\theta_{23}$ is
above 1/2, but values both below and above are preferred.

\begin{table*}[ht]
\caption{Values of the parameters for each case.}
\label{tab:cases}
\begin{ruledtabular}
\begin{tabular}{lccccc} 
 & $O_{23}^L$ & $O_{12}^L$ & $O_{23}^R$ & $O_{13}^R$ & $O_{12}^R$ \\[1mm] \hline 
I-A & -$\pi/4$ & $\sin^{-1} (1/\sqrt{3})$ & $\varepsilon\; t_{23}^R$ & $\varepsilon\; t_{13}^R$ & $\varepsilon\; t_{12}^R$ \\[1mm]
I-B & -$\pi/4$ & $\varepsilon\; t_{12}^L$ & $\varepsilon\; t_{23}^R$ & $\varepsilon\; t_{13}^R$ & $\sin^{-1} (1/\sqrt{3})$ \\[1mm]
I-C & -$\pi/4$ & $\sin^{-1} (1/2)$ & $\varepsilon\; t_{23}^R$ & $\varepsilon\; t_{13}^R$ & $\sin^{-1} (1/\sqrt{6})$ \\[1mm]
II-A & $\varepsilon\; t_{23}^L$ & $\varepsilon\; t_{12}^L$ & -$\pi/4$  & $\varepsilon\; t_{13}^R$ & $\sin^{-1} (1/\sqrt{3})$  \\[1mm]
II-B & $\sin^{-1} (1/\sqrt{3})$ & $\varepsilon\; t_{12}^L$ & -$\pi/4$  & $\varepsilon\; t_{13}^R$ & $\sin^{-1} ( 1/\sqrt{3})$ 
\\[1mm]  
\end{tabular}
\end{ruledtabular}
\end{table*}

\begin{table*}[t]
\caption{Mixing angles as function of the perturbed parameters $t_{ij}$.}
\label{tab:mixings}
\begin{ruledtabular}
\begin{tabular}{lccc}
 & $\vert V_{13} \vert^2$ & $sin^2 \theta_{atm}$ & $sin^2 \theta_{sol}$ \\[1mm] \hline 
I-A & $\frac{\varepsilon^2}{3} \left( 2 (t_{13}^R)^2 + (t_{23}^R)^2 \right)$ & $\frac{1}{2} - \frac{\varepsilon}{\sqrt{3}}\left( t_{13}^R \cos \alpha- \sqrt{2} t_{23}^R \sin \alpha \right)$ & $\frac{1}{3} + \frac{\varepsilon^2}{9} \left( 3 (t_{12}^R)^2 + 2(t_{13}^R)^2 -2 (t_{23}^R)^2 \right)$ \\[2mm]
I-B & $\varepsilon^2 (t_{13}^R)^2$ & $\frac{1}{2} +\varepsilon\,t_{23}^R \sin \alpha - \varepsilon^2 t_{12}^L t_{13}^R \cos \alpha $ & $\frac{1}{3} + \frac{\varepsilon^2 (t_{12}^L)^2}{3}$ \\[2mm]
I-C & $\frac{\varepsilon^2}{4} \left( 3 (t_{13}^R)^2 + (t_{23}^R)^2 \right)$ & $\frac{1}{2} - \frac{\varepsilon}{2}\left( t_{13}^R \cos \alpha+ \sqrt{3} t_{23}^R \sin \alpha \right)$ & $\frac{1}{3} + \frac{\varepsilon^2}{24} \left( 3 (t_{13}^R)^2 - 2 \sqrt{5} t_{13}^R t_{23}^R -3 (t_{23}^R)^2 \right)$\\[2mm]
II-A & $\frac{\varepsilon^2}{2} \left( (t_{12}^L)^2 + 2(t_{13}^R)^2 \right)$ & $\frac{1}{2} + \varepsilon\,t_{23}^L \sin \alpha- \frac{\varepsilon^2 (t_{12}^L)^2}{4}$ & $\frac{1}{3} + \frac{\varepsilon^2 (t_{12}^L)^2}{6}$\\[2mm]
II-B & $\frac{\varepsilon^2}{2} \left( (t_{12}^L)^2 + 2(t_{13}^R)^2 \right)$ & $\frac{1}{2}+ \frac{\sqrt{2} \sin \alpha}{3} -\frac{\varepsilon^2}{12} \left( (t_{12}^L)^2 - 8 t_{12}^L t_{13}^R \cos \alpha \right)$ & $\frac{1}{3} + \frac{\varepsilon^2 (t_{12}^L)^2}{6}$ \\[1mm] 
\end{tabular}
\end{ruledtabular}
\end{table*}

Given the closeness of tribimaximal mixing with experimental values, we fix
some of the parameters such that we can reproduce TBM to zeroth order. The
remaining parameters are then small and can be treated as perturbation
parameters $\theta_{ij}=\varepsilon\,t_{ij}$, with $\varepsilon$ of the
order the Cabibbo angle. We identify five different cases. In 
Table~\ref{tab:cases}, we show the values for the parameters being used in our 5
different cases. Table~\ref{tab:mixings} shows the explicit expression for
the mixing angles, in terms of the perturbation parameters $\varepsilon
\,t_{ij}$ for each of the cases. All cases can have large Dirac CP violation.

--\textbf{Scenario I-A}:

This scenario yields in leading order a value for the Dirac-type invariant 
$I_{CP}$, which may be large: 
\begin{equation}
\begin{aligned} I_{CP} =\, \frac{\varepsilon}{6 \sqrt{3}} \left\vert
\sqrt{2} t_{23}^R \cos \alpha - 2 t_{13}^R \sin \alpha \right\vert\,,
\end{aligned}  \label{jcp1a}
\end{equation}

All experimental results on mixing, including the central value for the
solar angle, can be fit with just the phase $\alpha$, and the small
parameter combination $\varepsilon \,t_{23}^{R}$ of the order of the Cabibbo
angle. If we take the limit of small $t_{12}^{R}$ and $t_{13}^{R}$, a non
zero value for $\alpha$ is necessary to have a value of $\sin \theta
_{atm}\neq 1/2$. In addition, if the $t_{12}^{R}$ and $t_{13}^{R}$ are
small, the Majorana-CP violating phases are large ($\sim \pi /2$). We find
for the Majorana phases: 
\begin{equation}
\tan \alpha_{1}^{M}\,=\,\frac{\sqrt{2}}{\varepsilon \,t_{12}^{R}}\,,\qquad
\tan \alpha_{2}^{M}\,=\,\frac{t_{23}^{R}}{\sqrt{2}\,t_{13}^{R}}\,.
\end{equation}
Clearly, the Majorana phases will decrease if $(t_{12}^{R},\,t_{13}^{R})$
assume substantial values, but that will increase the value for the solar
angle. We find for the Double-Beta Decay parameter, (the leading order
approximation) for the quasi-degenerate case, 
\begin{equation}
M_{ee}\,=\,\frac{\mu}{3}\,,
\end{equation}
Another important aspect of this scenario is the form the neutrino mass
matrix for the quasi-degenerate case. In leading order, we find: 
\begin{equation}
M\,=\,\frac{\mu}{3}
\begin{pmatrix}
1 & -2 & -2 \\ 
-2 & \frac{-1+3e^{-2i\alpha}}{2} & \frac{1+3e^{-2i\alpha}}{2} \\ 
-2 & \frac{1+3e^{-2i\alpha}}{2} & \frac{-1+3e^{-2i\alpha}}{2}
\end{pmatrix}
\,.  \label{mvtbm}
\end{equation}

Furthermore, we obtain for the CP violation quantity $G$, defined in 
Eq.~\eqref{g3}: 
\begin{equation}
G\,=\,\frac{4}{9}\left\vert \sin (2\alpha )\right\vert \,.  \label{g1a}
\end{equation}

--\textbf{Scenario I-B}: The CP-Invariant is in this case (in leading order)
given by 
\begin{equation}
I_{CP}\,=\,\frac{\varepsilon}{3\sqrt{2}}\lvert t_{13}^{R}\cos \alpha \rvert
\,,
\end{equation}
If we want to avoid the central value for the atmospheric mixing angle,
then, it is clear that we need at least 3 parameters, $\alpha
,t_{13}^{R},t_{23}^{R}$, to fit the experimental results on mixing and large
Dirac-CP violation. The central value for the solar angle can not be
achieved, not even with the use of all parameters. The Majorana-CP violating
phases are 
\begin{equation}
\tan \alpha_{1}^{M}\,=\,\frac{3}{\sqrt{2}}\varepsilon \,t_{12}^{L}\,,\quad
\tan \alpha_{2}^{M}\,=\,\frac{\varepsilon \,t_{12}^{L}}{\sqrt{2}}\left( \,1+
\frac{\sqrt{2}t_{23}^{R}}{t_{13}^{R}}\right) .
\end{equation}
This scenario produces small Majorana-CP violating phases. If neutrinos are
quasi-degenerate, we find for the neutrino mass matrix, in leading order 
\begin{equation}
M\,=\,\mu 
\begin{pmatrix}
1 & 0 & 0 \\ 
0 & \sin \alpha & \cos \alpha \\ 
0 & \cos \alpha & -\sin \alpha
\end{pmatrix}
\,.  \label{mv1b}
\end{equation}
The Double-Beta Decay parameter and the CP violation quantity $G$ read (in
leading order): 
\begin{equation}
M_{ee}\,=\,\mu \quad \text{and}\quad G\,=\,0\,,
\end{equation}
respectively.

--\textbf{Scenario I-C}: An intermediate scenario where both $O_{12}^{L}$
and $O_{12}^{R}$ are large. We choose one of the many combinations of these
two angles to obtain TBM mixing. Then, three parameters are fixed, which
leaves only 3 free parameters.This scenario yields for $I_{CP}$, 
\begin{equation}
\begin{split}
I_{CP}\,=\,& \frac{\varepsilon}{24}\left\vert \left( \sqrt{15}t_{13}^{R}+
\sqrt{3}t_{23}^{R}\right) \cos \alpha \right. \\
& \left. +\left( \sqrt{5}t_{23}^{R}-3t_{13}^{R}\right) \sin \alpha
\right\vert \,.
\end{split}
\end{equation}
Again, we may have large Dirac-CP violation. If we want to avoid the central
value for the atmospheric mixing angle, it may be seen here, that we only
need 2 parameters: the phase $\alpha$, and one of the remaining $t_{ij}$,
to fit the experimental results on mixing, but remember that this depends on
the choice of the two large angles of $O_{12}^{L}$ and $O_{12}^{R}$. In this
context, the Majorana-CP violating phases can be large: 
\begin{equation}
\tan \alpha_{1}^{M}\,=\,3\sqrt{\frac{3}{5}}\,,\quad \tan \alpha_{2}^{M}\,=\,
\frac{\sqrt{3}\left( t_{13}^{R}+\sqrt{5}t_{23}^{R}\right)}{3\sqrt{5}
t_{13}^{R}-t_{23}^{R}}\,.
\end{equation}
We find for the quasi-degenerate limit the neutrino mass matrix, the
Double-Beta Decay parameter and the CP violation quantity $G$, in leading
order: 
\begin{equation}
M\,=\,\frac{\mu}{2}
\begin{pmatrix}
1 & -\sqrt{\frac{3}{2}} & \sqrt{\frac{3}{2}} \\ 
-\sqrt{\frac{3}{2}} & \frac{2e^{-2i\alpha}-1}{2} & \frac{2e^{-2i\alpha}+1}{2}
\\ 
\sqrt{\frac{3}{2}} & \frac{e^{-2i\alpha}+1}{2} & \frac{e^{-2i\alpha}-1}{2}
\end{pmatrix}
\,,
\end{equation}
\begin{equation}
M_{ee}\,=\frac{\mu}{2}\,,\quad G\,=\,\frac{9}{16}\left\vert \sin 2\alpha
\right\vert \,.  \label{meeg}
\end{equation}

--\textbf{Limit case II}: As in Limit case I, we may construct here two
opposite and distinctive scenarios: a scenario where $O_{23}^{L}$ is large,
or a scenario where $O_{23}^{R}$ is large. The scenario where $O_{23}^{L}$
is large, but where $O_{23}^{R}$ is small, is already contained in the
scenario I-A of Limit case I (modulo some slight modifications which produce
equivalent results). It is therefore sufficient to focus on a scenario where 
$O_{23}^{L}$ is small and $O_{23}^{R}$ is large, or exceptionally on a
scenario between, where both are large.

--\textbf{Scenario II-A}: a scenario where $O_{23}^{L}$ is small and 
$O_{23}^{R}$ is large. The Dirac CP invariant is given (in leading order) by 
\begin{equation}
I_{CP}\,=\,\frac{\varepsilon}{6}\,|t_{12}^{L}|\,.
\end{equation}

In this case, it is clear that we cannot achieve a central value for the
solar angle, we need $t_{12}^{L}\neq 0$ to have a non zero value for 
$I_{CP}$, but doing so will increase the value of the solar angle above one sigma.
The Majorana-CP violating phases are also obtained in leading order 
\begin{equation}
\tan \alpha_{1}^{M}\,=\,\frac{3}{2}\varepsilon \,t_{12}^{R}\,,\qquad \tan
\alpha_{2}^{M}\,=\,\frac{t_{12}^{L}}{\sqrt{2}t_{13}^{R}}\,,
\end{equation}
where only the second one can be large. For quasi-degenerate neutrinos, we
find for the neutrino mass matrix, the Double-Beta Decay parameter and the
CP violation quantity $G$, in leading order: 
\begin{equation}
M\,=\,\mu \ {1\>\!\!\!\mathrm{I}}\,,\quad M_{ee}\,=\,\mu \,,\quad G=0\,.
\end{equation}

--\textbf{Scenario II-B}: an intermediate scenario where both $O_{23}^{L}$
and $O_{23}^{R}$ are large.

Also for this case only 2 parameters are needed, e.g., the perturbative
parameter $t_{13}^{R}$ and the phase $\alpha$, which has to be small of the
order of the Cabibbo angle, to fit the experimental results on atmospheric
mixing and large Dirac-CP violation, but again here this depends on the
choice for the two large angles of $O_{23}^{L}$and $O_{23}^{R}$. In first
order, we have for the Dirac CP invariant: 
\begin{equation}
I_{CP}\,=\,\frac{\varepsilon}{18}\left\vert t_{12}^{L}+4t_{13}^{R}\cos
\alpha \right\vert \,.
\end{equation}
As in all of the previous cases, one can have a large value for $I_{CP}$.
For this case, we can make a simple (leading order) prediction if we take 
$t_{12}^{L}$ much smaller than $t_{13}^{R}$ and $\alpha $ small:
\begin{equation}
I_{CP}=\frac{2}{9}\left\vert V_{13}\right\vert  \label{pr}
\end{equation}
The Majorana-CP violating phases are obtained in leading order 
\begin{equation}
\tan \alpha_{1}^{M}\,=\,\frac{3}{2}\varepsilon \,t_{12}^{R}\,,\qquad \tan
\alpha_{2}^{M}\,=\,\frac{t_{12}^{L}}{\sqrt{2}t_{13}^{R}}\,.
\end{equation}
where again, the second one can be large. For quasi-degenerate neutrinos, we
find for the neutrino mass matrix, the Double-Beta Decay parameter and the
CP violation quantity $G$, in leading order: 
\begin{equation}
M\,=\,\mu 
\begin{pmatrix}
1 & 0 & 0 \\ 
0 & \frac{e^{-2i\alpha}-2}{3} & \frac{\sqrt{2}(1+e^{-2i\alpha})}{3} \\ 
0 & \frac{\sqrt{2}(1+e^{-2i\alpha})}{3} & \frac{2e^{-2i\alpha}-1}{3}
\end{pmatrix}
\,,  \label{meeg3}
\end{equation}
\begin{equation}
M_{ee}\,=\,\mu \,,\quad G=0\,.
\end{equation}
As already mentioned in this section, it can be seen from 
Table~\ref{tab:mixings} that in the cases I-B, II-A and II-B the value for 
$\sin^{2}\theta_{sol}$ can not be lower than $1/3$, which is not in agreement
with the experimental results given Eq.~\eqref{tortola} at one-sigma level.
This is of course due to our initial choice in Eq.~\eqref{iTBM}, which
corresponds to exact tribimaximal mixing. We stress that some of our
conclusions with regard to the different scenarios may depend significantly
on the initial starting point, while others do not. However, with regard to
Scenario I-A, very similar results are obtained if one chooses as starting
points e.g. the golden ratio mixing of type I~\cite{Kajiyama:2007gx} or the
hexagonal mixing~\cite{Giunti:2002sr,Xing:2002az}, instead of TBM.

\textbf{Scenario I-A and the Standard Parametrization} We are tempted to
find Scenario I-A the most appealing. It only needs 2 extra parameters to
fit the experimental results on lepton mixing and provides large Dirac-CP
violation and large values for the Majorana-CP violating phases. The other
scenarios need more parameters, or need more adjustment. We also point out
that Scenario I-A, would not appear so clearly, if one used a different
parametrization, e.g. one the parametrizations mentioned just after 
Eq.~\eqref{pmns1}.

\begin{figure}[t]
\includegraphics[width=0.5\textwidth]{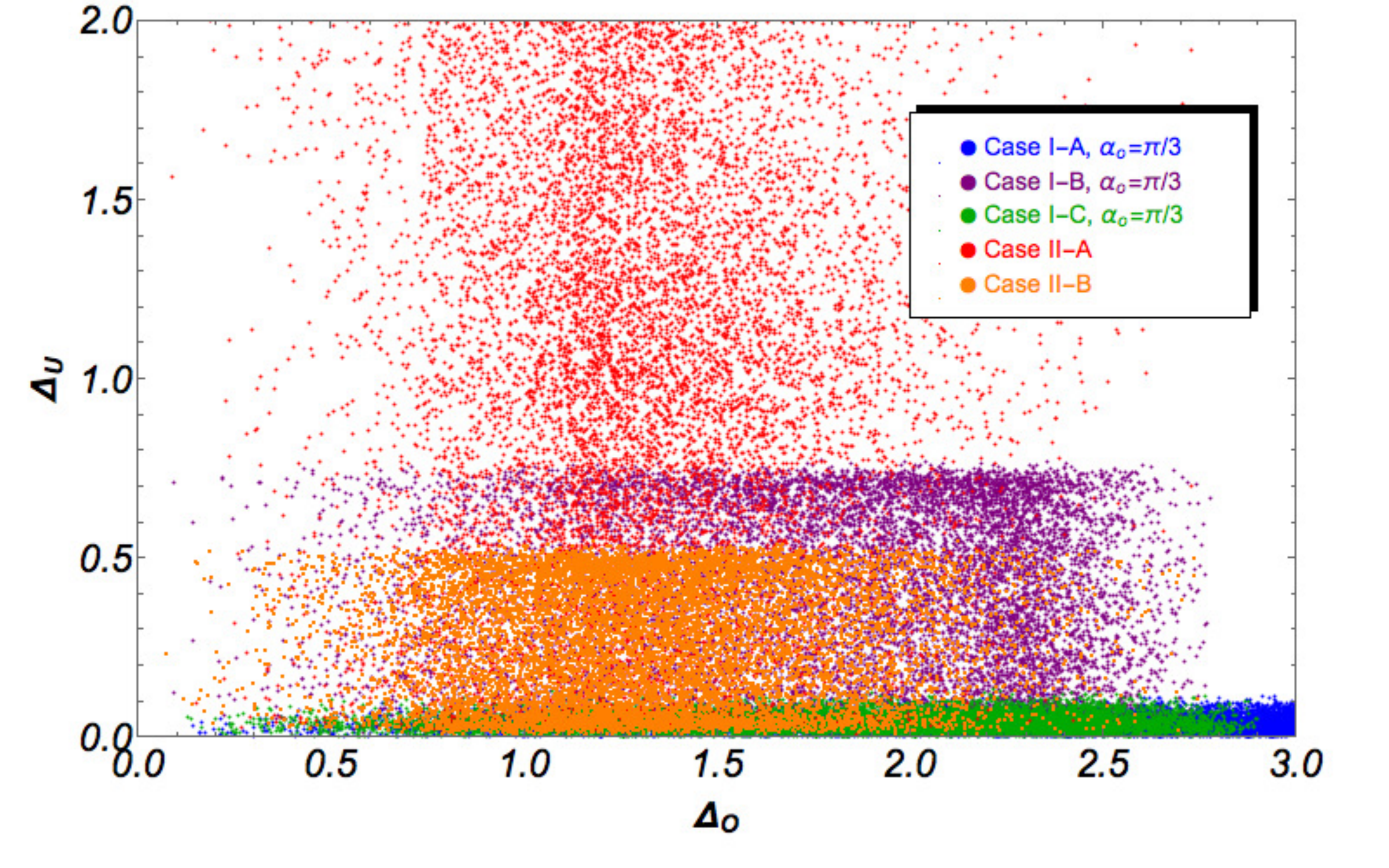}
\caption{Plotting $\Delta_O$ versus $\Delta_U$ for the five cases
indentified in Eq.~\eqref{tbm1} with $\protect\alpha_{o}=\protect\pi/3$.}
\label{fig:DeltaODeltaU}
\end{figure}

\begin{figure}[t]
\includegraphics[width=0.5\textwidth]{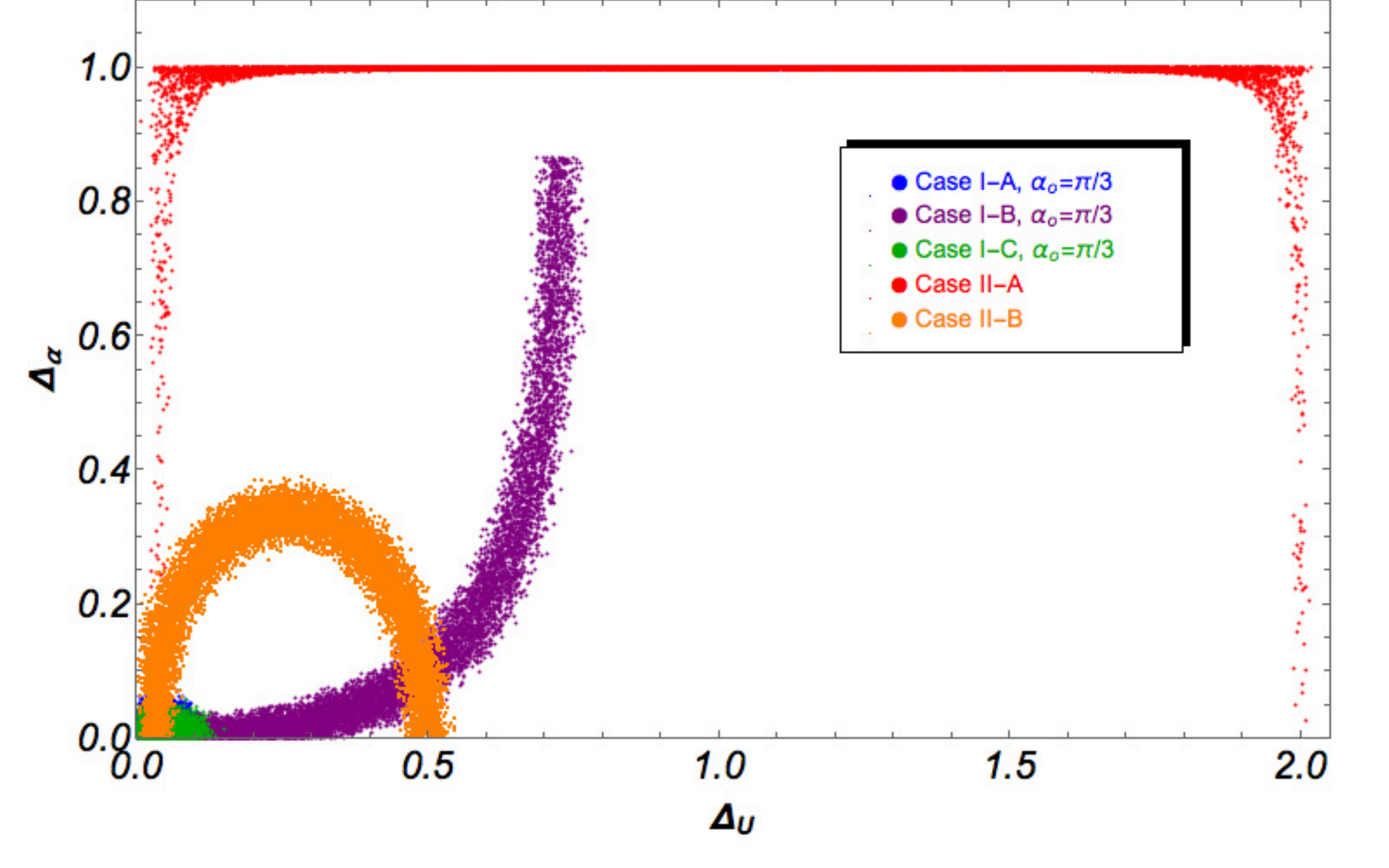}
\caption{Plotting $\Delta_O$ versus $\Delta_U$ for the five cases
indentified in Eq.~\eqref{tbm1} with $\protect\alpha_{o}=\protect\pi/3$.}
\label{fig:DeltaUDeltaAlpha}
\end{figure}

\begin{figure}[t]
\includegraphics[width=0.5\textwidth]{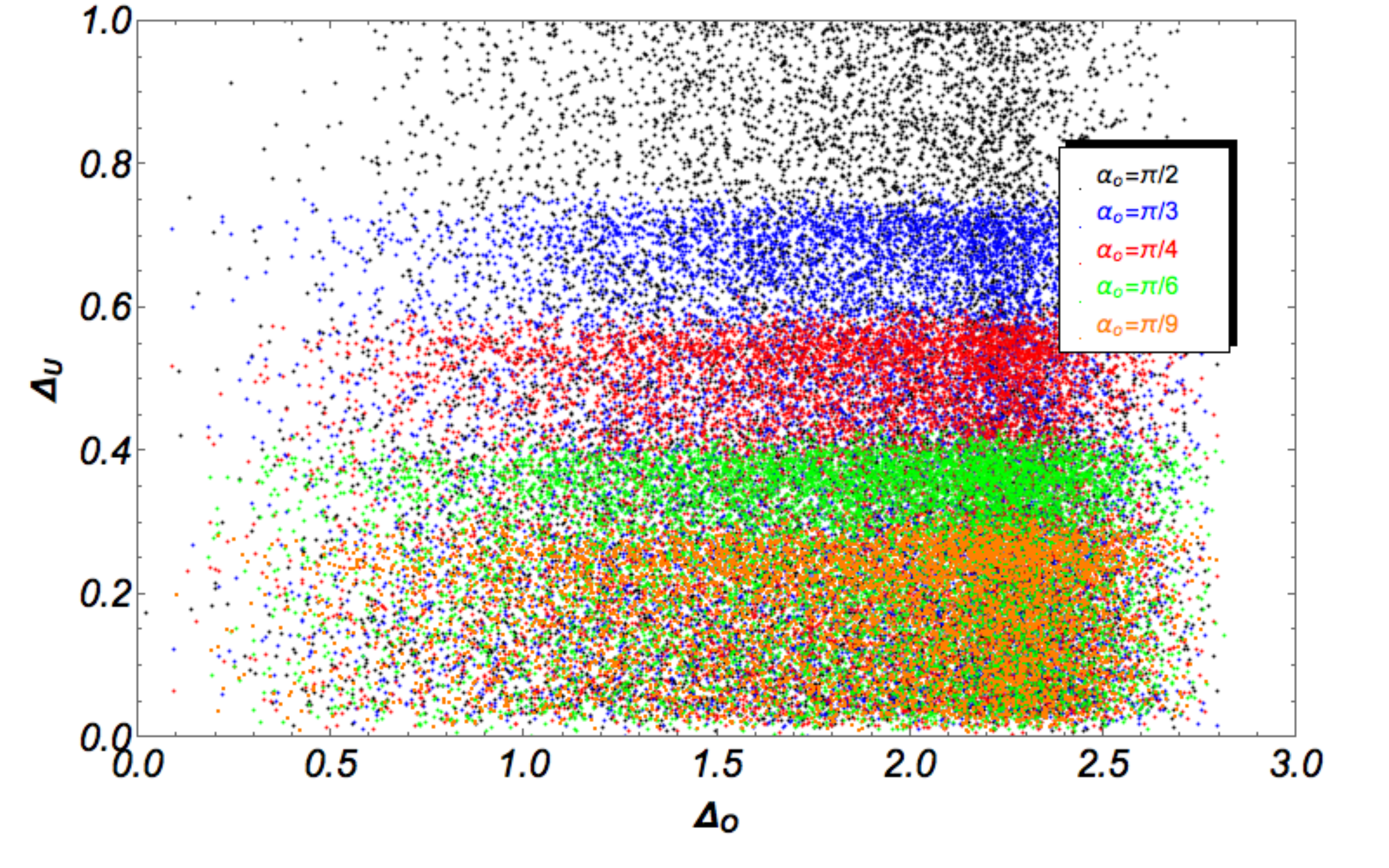}
\caption{Plotting $\Delta_O$ versus $\Delta_U$ for the case I-B varying 
$\protect\alpha_{o}=\protect\pi/2$, $\protect\pi/3$, $\protect\pi/4$, 
$\protect\pi/6$ and $\protect\pi/9$.}
\label{fig:1BDeltaODeltaU}
\end{figure}

Given the relevance of Scenario I-A, we shall now reproduce this scenario in
the Standard Parametrization given in Eq.~\eqref{eq:spParm}, where the
TBM-scheme is obtained with 
\begin{equation}
V^{SP}\,=\,O_{23}^{\pi /4}\cdot K_{_{D}}\cdot O_{13}\cdot O_{12}^{\phi
_{o}}\,,\quad \sin \phi_{o}=\frac{1}{\sqrt{3}}\,,  \label{1asp}
\end{equation}
with the angle of $O_{13}$ put to zero. For simplicity, we leave out the
Majorana phases. In this parametrization, in order to have a value for 
$\left\vert V_{13}\right\vert \neq 0$, we have to switch on the angle 
$O_{13}$. However, for the unitary matrix in Eq.~\eqref{1asp}, one may check that
even then, $\left\vert V_{23}\right\vert =\left\vert V_{33}\right\vert$,
irrespective of the value of the angle of $O_{13}$. Thus, using this
remaining parameter, one can not adjust the atmospheric mixing angle,
unless, e.g. from the start, the angle of the $O_{23}$ is chosen to be
different from $\pi /4$. One has to correct the atmospheric mixing angle, or
from the beginning, or afterwards, with some additional extra contribution
which modifies the TBM-limit. It is clear, in the Standard Parametrization,
adjusting the TBM-limit for the atmospheric mixing angle is not possible
using the remaining parameters. This is in clear contrast with our new
parametrization and what we obtain for Scenario I-A, where the parameters
available in the actual parametrization, in this case, via suitable choice
for of the parameter $\varepsilon \,t_{23}$ in Eq.~\eqref{1asp}), at the
same time adjust the atmospheric mixing angle, generate a small value for 
$\left\vert V_{13}\right\vert$, and make possible large values for CP
violation. Possibly, this may be useful for some model.

\begin{figure*}[t]
\centering
\includegraphics[width=0.47\textwidth]{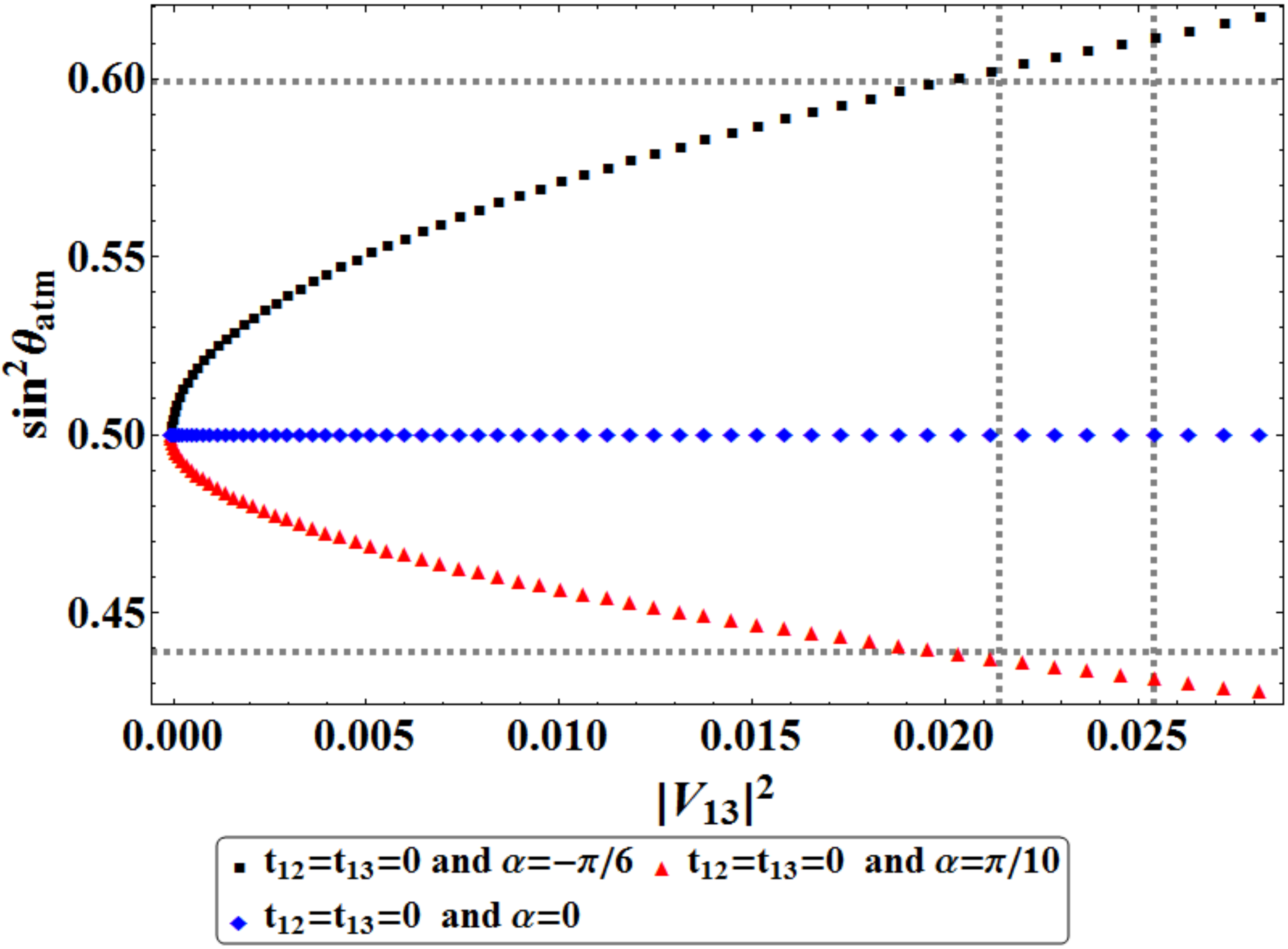} \hfill 
\includegraphics[width=0.47\textwidth]{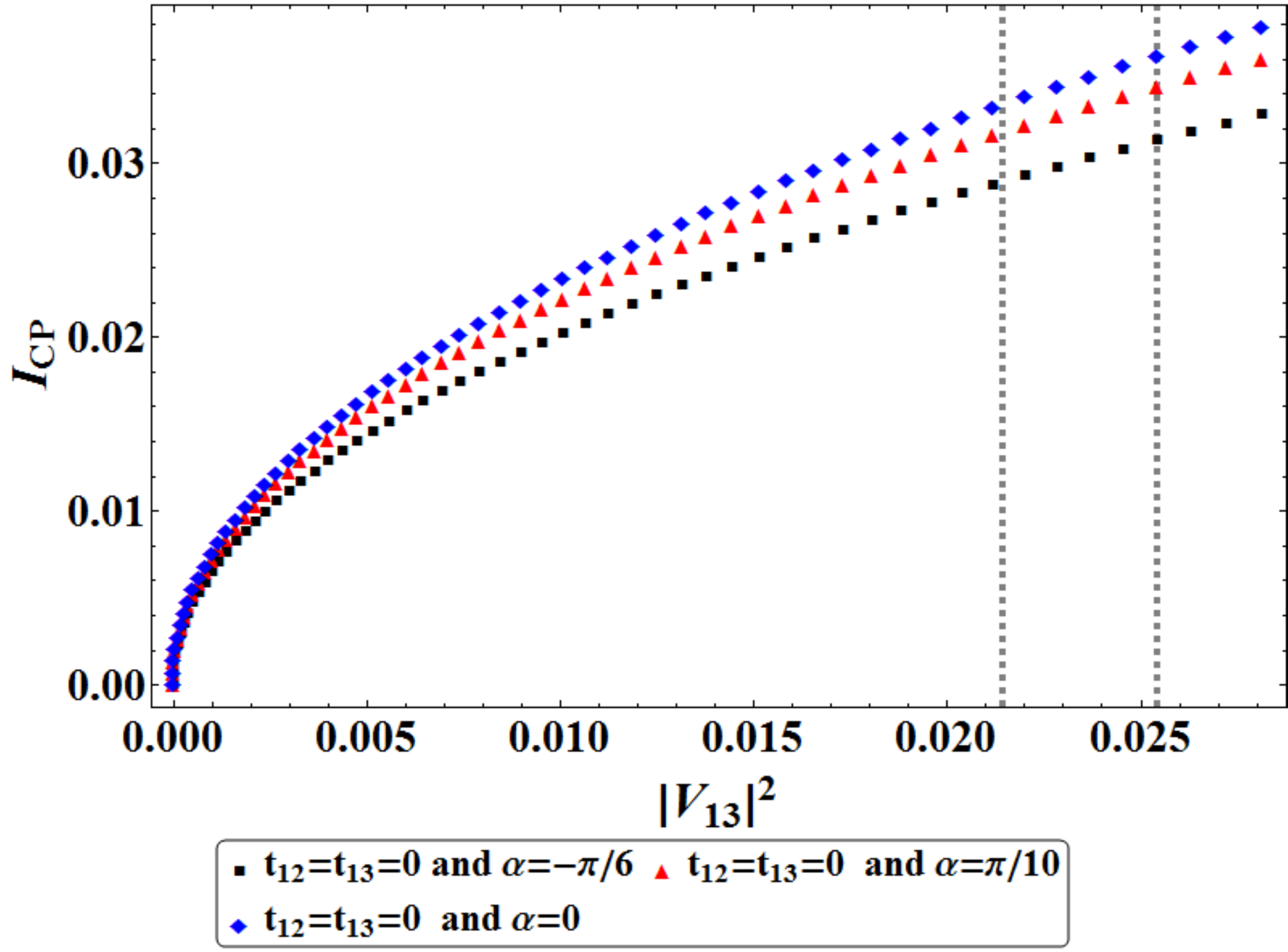}
\caption{Plotting $\sin^2\protect\theta_{atm}$ and the CP invariant $I_{cp}$
as a function of $|V_{13}^2|$ for scenario I-A.}
\label{fig:IcpIA}
\end{figure*}

\begin{figure*}[t]
\centering
\includegraphics[width=0.47\textwidth]{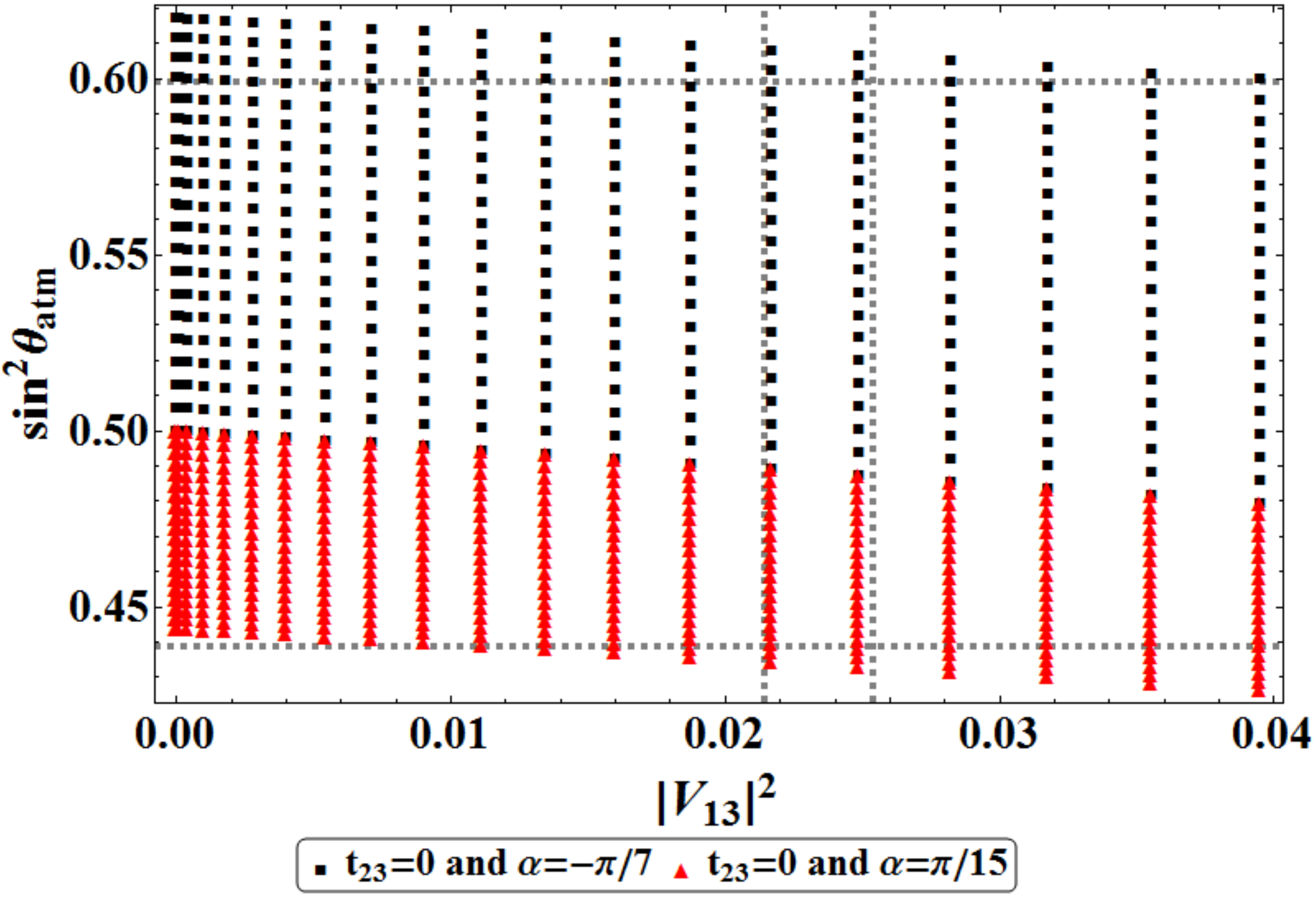} \hfill
\includegraphics[width=0.47\textwidth]{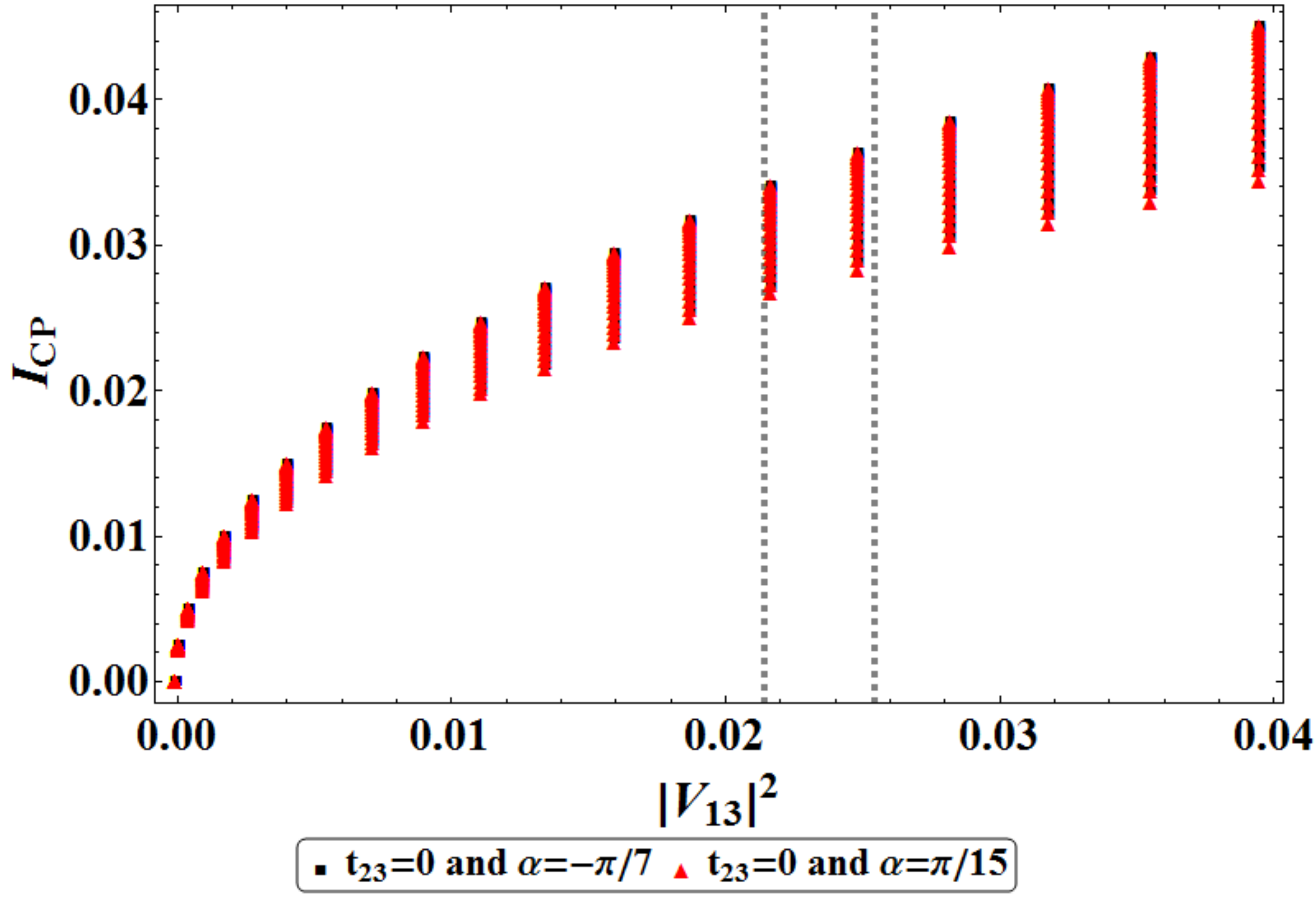}
\caption{Plotting $\sin^2\protect\theta_{atm}$ and the CP invariant $I_{cp}$
as a function of $|V_{13}^2|$ for scenario II-A.}
\label{fig:IcpIIA}
\end{figure*}

\begin{figure*}[t]
\centering
\includegraphics[width=0.47\textwidth]{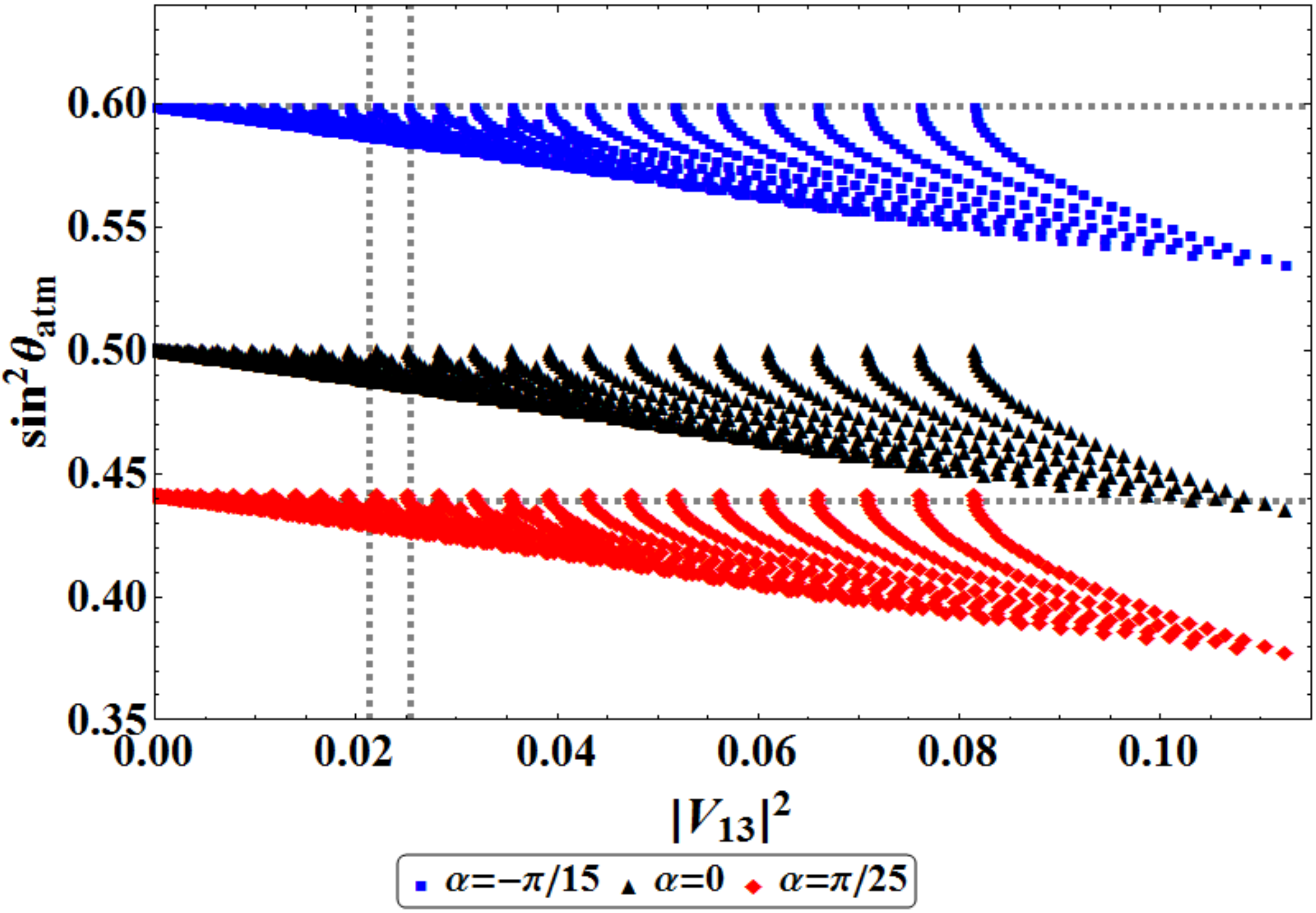} \hfill 
\includegraphics[width=0.47\textwidth]{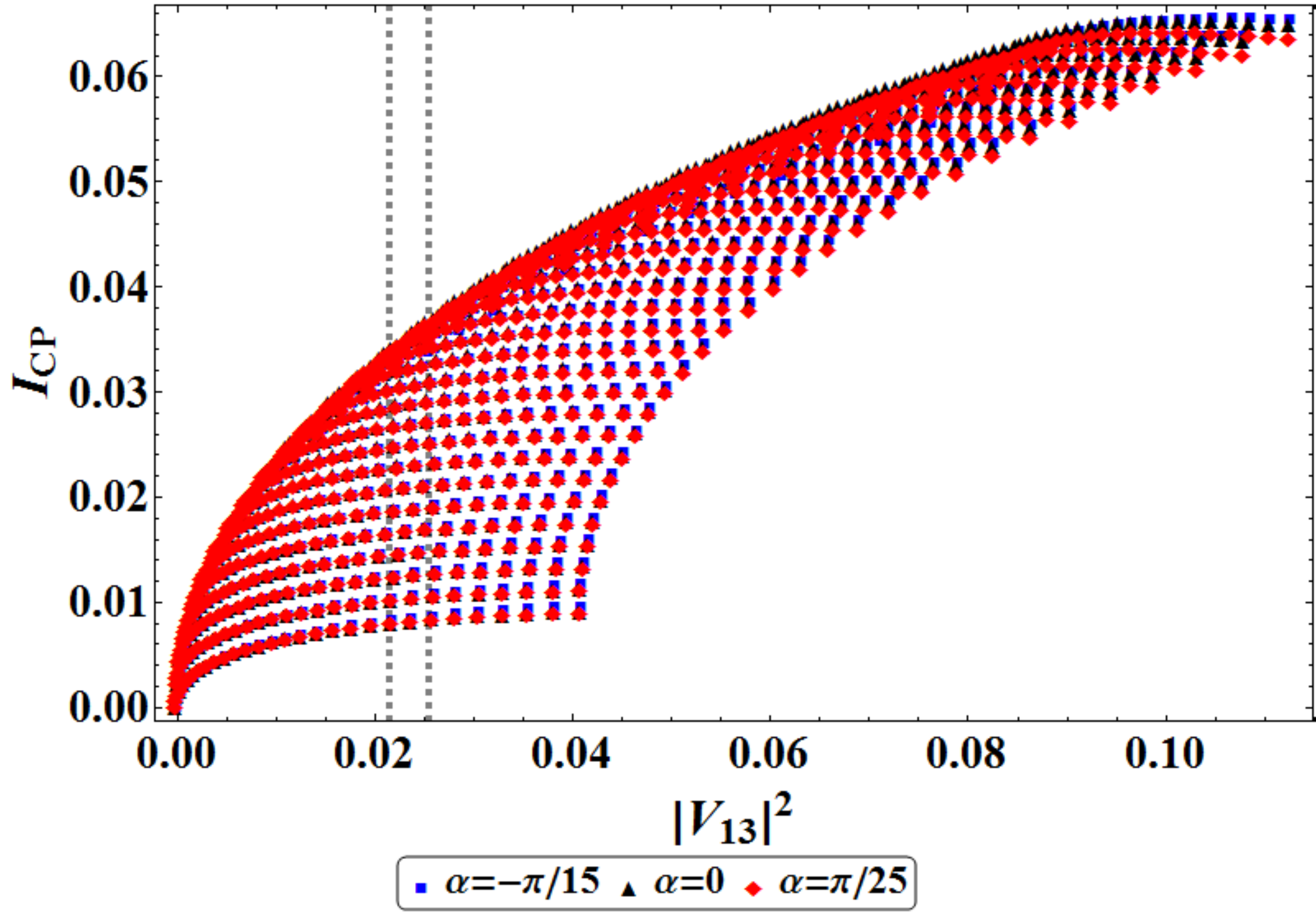}
\caption{Plotting $\sin^2\protect\theta_{atm}$ and the CP invariant $I_{cp}$
as a function of $|V_{13}^2|$ for scenario II-B.}
\label{fig:IcpIIB}
\end{figure*}

\section{\label{num} Numerical Simulation and Stability}

For completeness, we give a numerical analysis of some of the scenarios
described in the previous section. We choose a fixed scheme, the TBM scheme
constructed with the 5 different scenarios. More precisely, we test 
\begin{equation}  \label{tbm1}
\begin{aligned} \text{I-A} &:\qquad V_{o}=O_{23}^{\pi
/4}O_{12}^{\phi_{o}}\,K_{\alpha_{o}}^{i} \\ \text{I-B} &: \qquad
V_{o}=O_{23}^{\pi/4}\,K_{\alpha_{o}}^{i}O_{12}^{\phi_{o}} \\ \text{I-C} &:
\qquad V_{o}=O_{23}^{\pi
/4}\,O_{12}^{\phi_1}K_{\alpha_{o}}^{i}O_{12}^{\phi_{2}} \\ \text{II-A} &:
\qquad V_{o}=O_{23}^{\pi/4}O_{12}^{\phi_{o}}\, \\ \text{II-B} &: \qquad
V_{o}=O_{23}^{\theta_{o}}\,K_{\alpha_{o}}^{i}O_{23}^{\pi
/4}O_{12}^{\phi_{o}} \end{aligned}
\end{equation}
where $\sin \phi_{o}=\sin \theta_{o}=\frac{1}{\sqrt{3}}$, $\sin \phi_{1}= 
\frac{1}{2}$, $\sin \phi_{2}=\frac{1}{\sqrt{6}}$. We define the $U_{o}$ as
the matrix on the left of, together with the $K_{\alpha_{o}}^{i}$. In the
II-A case, this is the identity matrix. We define also the $O_{o}$ as the
matrix on the right of the $K_{\alpha_{o}}^{i}$. In the II-A case, this is
the whole matrix $V_{o}$. For case II-B, $\alpha_{o}=0$ as pointed out in
the previous section. For the other cases, we assume for $\alpha_{o}$,
diverse fixed values.

We illustrate in Figs.~\ref{fig:IcpIA}-\ref{fig:IcpIIB} the correlations
among the observables for the scenarios I-A, II-A and II-B. The figures plot
for each scenario $\sin ^{2}\theta_{atm}$ and $I_{CP}$ as a function of 
$|V_{13}|^{2}$ and $I_{CP}$ as a function of $|V_{13}|^{2}$, for particular
values of the parameters left unconstrained in the definition of each
scenario according to Table~\ref{tab:cases}. Scenarios I-B and I-C are
omitted since they have similar behavior as scenario I-A for these
observables. A numerical analysis of Scenario I-A, was also done in 
Ref.~\cite{Branco:2014zza}. We can conclude from 
Figs.~\ref{fig:IcpIA}-\ref{fig:IcpIIB} that a large CP invariant 
$I_{CP}$ can be obtained in agreement
with the allowed experimental range of the observed parameters.

Next, we test how the lepton mixing matrix changes and the stability of our
scenarios, by adding a small random perturbation to a predefined exact
degenerate limit. To do this, we construct a neutrino mass matrix $M$,
composed of an exact degenerate part in the form of a symmetric unitary
matrix $S_{o}$ related to one of the TBM scenario schemes in 
Eq.~\eqref{tbm1}, and a part composed of a small random perturbation $Q^{\varepsilon}$.
Thus, the full quasi-degenerate neutrino mass matrix is as in Eq.~\eqref{mat1}: 
\begin{equation}
M\,=\,\mu \,\left( S_{o}+Q^{\varepsilon}\right) \,,  \label{mat2}
\end{equation}
where $S_{o}=U_{o}^{\ast}\,U_{o}^{\dagger}$, with the $U_{o}$'s of the
different cases, and $Q^{\varepsilon}$ is some small complex symmetric
random perturbation: 
\begin{equation*}
Q^{\varepsilon}\equiv \varepsilon ^{2}\ Q\,,\quad \varepsilon ^{2}=
\frac{\left( \Delta m_{31}^{2}\right) ^{\exp}}{2\mu ^{2}}\,,
\end{equation*}
with $Q$ random perturbations of $O(1)$ generated by our program. We test
the stability of lepton mixing of the different scenarios. We do not worry
about the exact mass differences, with two (reasonable) exceptions: we take
for $\varepsilon ^{2}\ $a fixed value. Inserting $\left( \Delta
m_{31}^{2}\right) ^{\exp}=2.5\times 10^{-3}\,\text{eV}^{2}$ together with a
common neutrino mass $\mu \gtrsim 0.15\,\text{eV,}$ we obtain $\varepsilon
^{2}\lesssim 0.05$; $\varepsilon$ is of the order of the Cabibbo angle.
These values make sure that we are in a mass range where the computed output 
$\Delta m_{31}^{2}=O(1)\times 10^{-3}\,\text{eV}^{2}$. We discard cases
generated by the perturbation where $\left\vert \Delta m_{31}^{2}\right\vert
<\left\vert \Delta m_{21}^{2}\right\vert$. Further, we do not impose any
other restrictions on the random perturbation $Q$ other than $Re(Q_{ij})$
and $Im(Q_{ij})$ to be real numbers between -1 and 1. However, we have
checked that further restrictions on the masses do not change significantly
any of the plots.

From the different mixing scenarios and the random $Q$'s in $M$, we compute
the full lepton mixing $V$, i.e. the corresponding diagonalizing matrix
matrix of $M$, such that $V^{\intercal}M\ V=D$ is real and positive.
Following the proof in Section~\ref{np}, we decompose the full lepton mixing 
$V$ in the new parametrization, obtained as in Eq.~\eqref{pmns}. We then
compare the new $U\equiv O_{23}\,O_{12}\,K_{\alpha}^{i}$ resulting from the
perturbation, with the original $U_{o}$ (i.e., without the perturbation) of
one of the cases in Eq.~\eqref{tbm1}, and evaluate a quantity $\Delta_{U}$
giving a reasonable measure of how much $U$ and $U_{o}$ differ: 
\begin{equation}
\Delta_{U}\,=\,\frac{1}{2}\sum \left\vert \,\left\vert U_{ij}\right\vert
-\left\vert (U_{o})_{ij}\right\vert \,\right\vert \,.  \label{y}
\end{equation}
Notice that this definition does not "see" the phase factors of the 
$K_{\alpha}^{i}$ of $U$, or of the $U_{o}$. For this, we evaluate the
changing on the phases $\alpha$ by defining the quantity 
\begin{equation}
\Delta_{\alpha}\,=\,\left\vert \,\left\vert \sin \alpha \right\vert
-\left\vert \sin \alpha_{o}\right\vert \,\right\vert \,,  \label{z}
\end{equation}
that compares the phase $\alpha$ of the $K_{\alpha}^{i}$ of $U$, with the
phase $\alpha_{o}$ of the $K_{\alpha_{o}}^{i}$ of $U_{o}$ and discarding
differences of $\pi $. The $II-A$ case, has no $\alpha_{o}$ phases. We have
also estimated how much $O$ in Eq.~\eqref{pmns} differs from our original 
$O_{o}$ in Eq.~\eqref{tbm1}, with 
\begin{equation}
\Delta_{O}\,=\,\frac{1}{2}\sum \left\vert \,\left\vert O_{ij}\right\vert
-\left\vert (O_{o})_{ij}\right\vert \,\right\vert \,,  \label{x}
\end{equation}
where again we discard any sign difference. The $1/2$ in front of $\Delta
_{O}$ (and $\Delta_{U}$) is a suitable normalization factor, chosen such
that, e.g. in a case where the original $O_{o}=\mathbbm{1}$ and the new $O$
is such that $O=O_{12}$ (or any other elementary rotation) with an angle 
$\sin\theta_{12}=0.2$, then also $\Delta_{O}\approx 0.2$, of the same
order of the Cabibbo angle.

In Fig.~\ref{fig:DeltaODeltaU} and ~\ref{fig:DeltaUDeltaAlpha} we plot 
$\Delta_{U}$ as a function of $\Delta_{O}$ and $\Delta_{\alpha}$ as a
function of $\Delta_{U}$, respectively, for the five scenarios. From 
Fig.~\ref{fig:DeltaODeltaU} and ~\ref{fig:DeltaUDeltaAlpha} we find that the 
$\Delta_U $ and $\Delta_\alpha$ of Scenarios I-A and I-C hardly suffer any
change with the perturbations. In Fig. 3, we show the variation of $\Delta_O$
as a function of $\Delta_U$ for different values of $\alpha_{o}=\pi/2$, 
$\pi/3$, $\pi/4$, $\pi/6$ and $\pi/9$ for Case I-B. Clearly, small $\alpha$
leads to more stability. Case I-A is not shown, since there is no apparent
change of these quantities by varying $\alpha$.

Cases I-A, I-C and I-B with small $\alpha$ are the most stable with regard
to $\Delta_U$ and $\Delta_\alpha$. As mentioned previously, Case I-C is
somewhat artificial as it requires a certain conspiracy between two angles 
$\phi_1$ and $\phi_2$ angles to be near the TBM limit. Therefore, we focus on
Case I-A. As shown in the previous section, generically, Scenario I-A has
also the largest Majorana phases.

\begin{figure*}[t]
\centering
\includegraphics[width=0.47\textwidth]{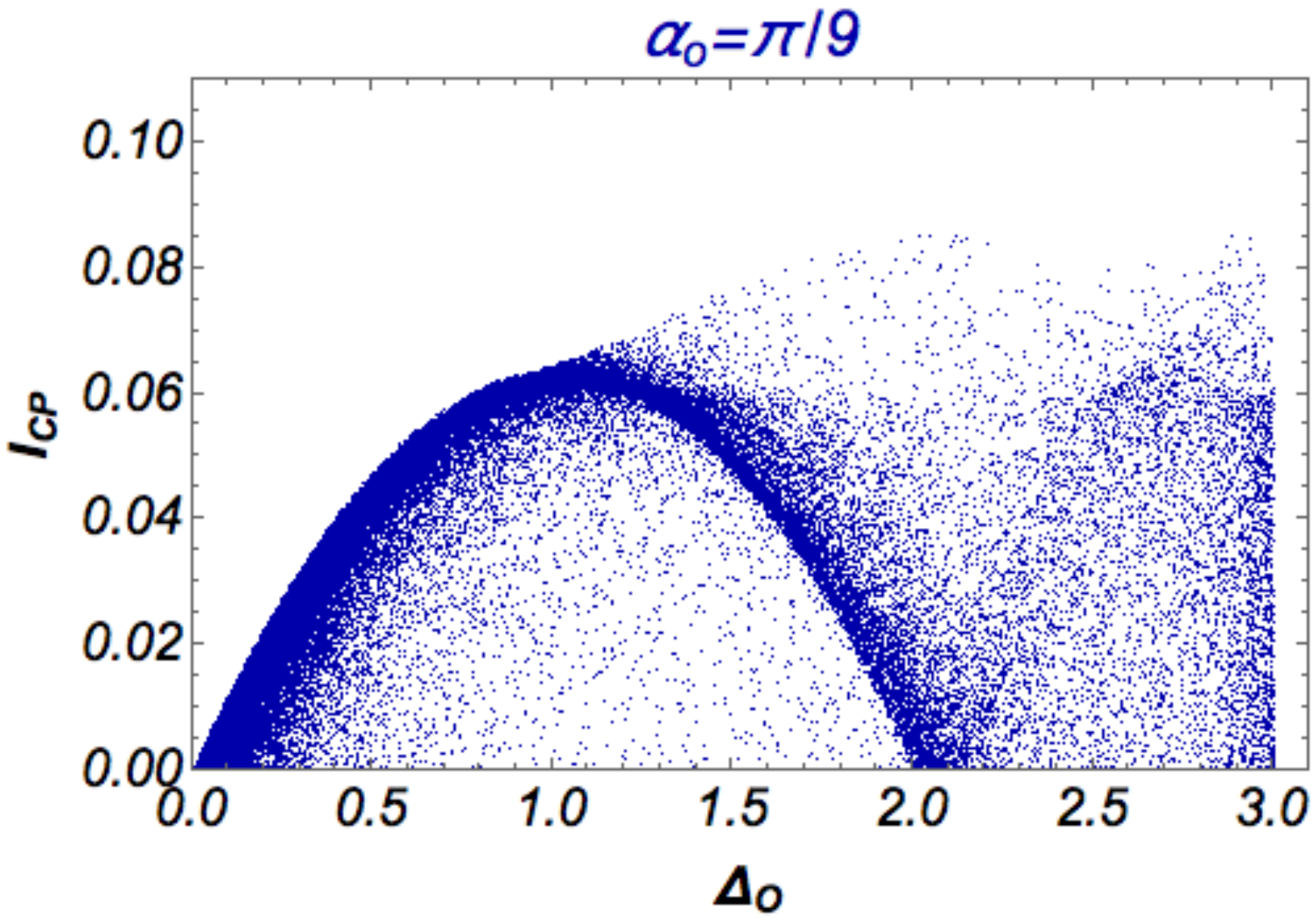} \hfill 
\includegraphics[width=0.47\textwidth]{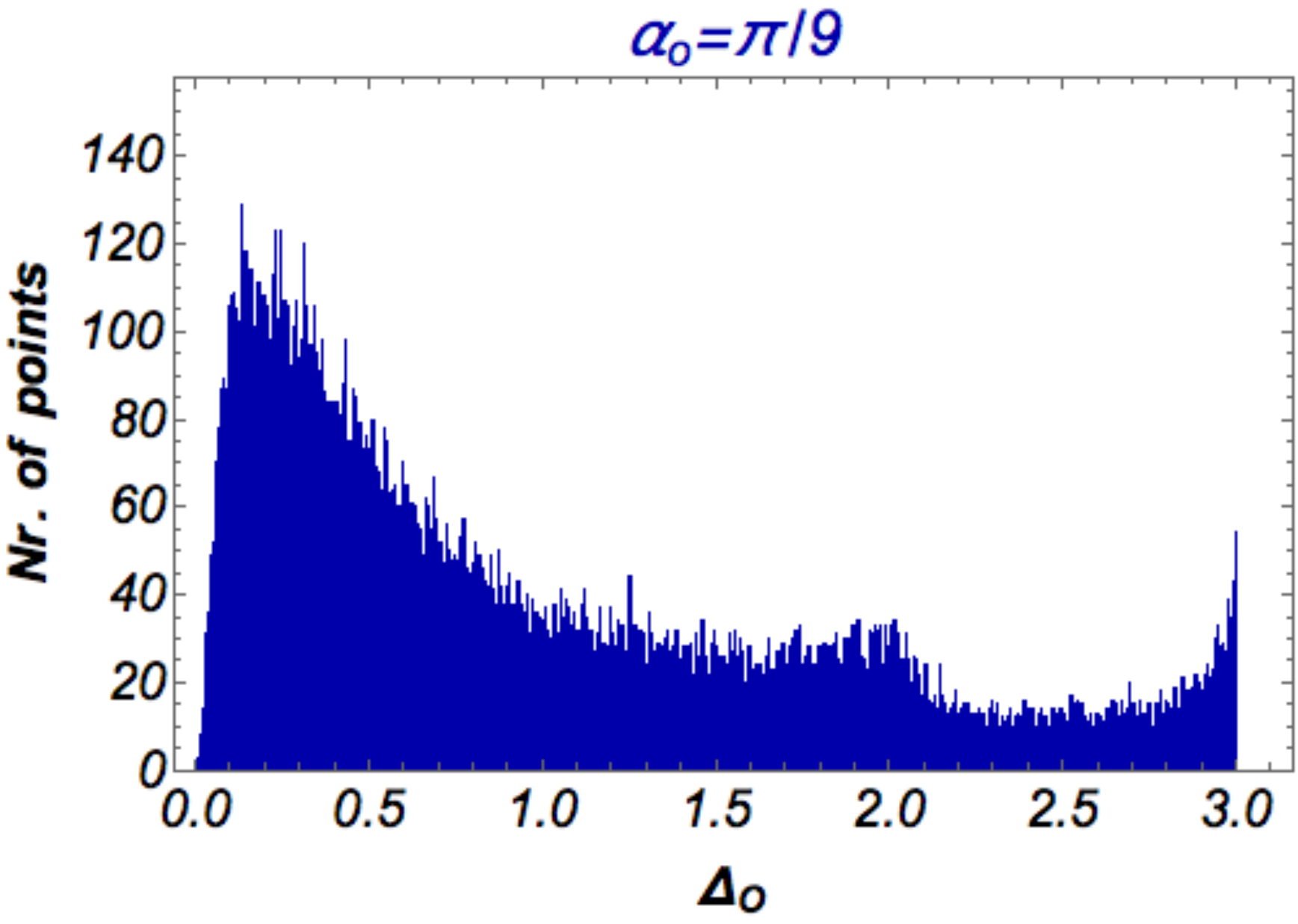}
\caption{\label{fig:QDeltaOvsIcp} Plotting the CP invariant $I_{CP}$ as a function of $\Delta_{O}$
considering restricted perturbations for $Q$. The right-handed plot is an
histogram showing the distribution of the values of $\Delta_{O}$ obtained
from $50000$ uniformly distributed sets of input values for $Q$, restricted
as indicated in Eq.~\eqref{qr}.}
\end{figure*}

With regard to the stability and variation of $\Delta_{O}$, we see that, in
general, the perturbations generate large $\Delta_{O}$ contributions for
all cases and in particular for Scenario I-A. It seems that this can only be
improved if one imposes restrictions on the allowed perturbations forcing
smaller $\Delta_{O}$'s. Maybe some kind of symmetry could accomplish this.
In Fig.~\ref{fig:QDeltaOvsIcp} we give an example, where the perturbations 
$Q $ are restricted: certain elements are taken to be zero, while the
imaginary part and the diagonal real part are taken to be $0.1$ smaller than
the others: 
\begin{equation}
Q\,=\,
\begin{pmatrix}
0 & 0 & 0 \\ 
0 & \frac{x_{22}}{10} & x_{23} \\ 
0 & x_{23} & \frac{x_{33}}{10}
\end{pmatrix}
+\frac{i}{10}
\begin{pmatrix}
0 & 0 & 0 \\ 
0 & y_{22} & y_{23} \\ 
0 & y_{23} & y_{33}
\end{pmatrix}
\label{qr}
\end{equation}
where the $x$'s, $y$'s are random real numbers varying between -1 and 1. For
the initial phase $\alpha_{o}$, we take $\alpha_{o}=\pi /9$. We see that
most of the deviations $\Delta_{O}$ (from the original $O_{o}=\mathbbm{1}$),
 are now around $0.2$ of the order of the Cabibbo angle, and this does not
affect having large values for $I_{CP}$.

\section{\label{sec:con} Conclusions}

We have studied some aspects of leptonic CP violation from a new
perspective. We have identified several limit scenario-cases, with mixing
angles in agreement with experiment and leading to large CP violation. We
proposed a new parametrization for leptonic mixing of the form 
$V=O_{23}\,O_{12}\,K_{\alpha}^{i}\cdot O$ to accomplish this.

If neutrinos are quasi-degenerate and Majorana, this parametrization is very
useful. It may reflect some specific nature of neutrinos, suggesting that
there is some major intrinsic Majorana character of neutrino mixing and CP
violation, present in the left part of the parametrization, while the
right part in the form of a real-orthogonal matrix $O$ with the 3 angles,
reflects the fact that there are 3 neutrino families with small mass
differences and results in small mixing. Thus, the intrinsic Majorana
character of neutrinos may be large with a large contribution to neutrino
mixing (from some yet unknown source), while the extra mixing $O$ of the
families is comparable to the quark sector and may be small, of the order of
the Cabibbo angle.

The new parametrization permits a new view of large leptonic CP violation.
It shows interesting aspects that were less clear for the standard
parametrization. We identified several limit scenario-cases and shown
results for mixing and CP violation. A certain scenario (I-A) was found to
be the most appealing. It only needs 2 extra parameters to fit the experimental
results on lepton mixing and provides large Dirac-CP violation and large
values for the Majorana-CP violating phases. We point out that the results
for this scenario derives explicitly from the form of the new
parametrization.

In addition and for quasi-degenerate Majorana neutrinos, the stability of
the different scenarios was tested using random perturbations. We concluded
that the left part of the parametrization behaves quite differently for the
diverse scenarios. Scenario I-A was very stable in this respect. With
respect to the right part of the parametrization, i.e., the real-orthogonal
matrix $O$, the perturbations generate large contributions for all cases.
This unstable part of the mixing (due to the random perturbations) can only
be improved, if one imposes restrictions on the allowed perturbations. We
have shown how to accomplish this.

\section*{Acknowledgments}

This work is partially supported by Funda\c{c}\~{a}o para a Ci\^{e}ncia e a
Tecnologia (FCT, Portugal) through the projects CERN/FP/123580/2011,
PTDC/FIS-NUC/0548/2012, EXPL/FIS-NUC/0460/2013, and CFTP-FCT Unit 777
(PEst-OE/FIS/UI0777/2013) which are partially funded through POCTI (FEDER),
COMPETE, QREN and EU. DW are presently supported by a postdoctoral
fellowship of the project CERN/FP/123580/2011 and his work is done at
CFTP-FCT Unit 777. D.E.C. thanks CERN Theoretical Physics Unit for
hospitality and financial support. D.E.C.~was supported by 
Associa\c{c}\~{a}o do Instituto Superior T\'{e}cnico para a 
Investiga\c{c}\~{a}o e Desenvolvimento (IST-ID).

\bibliography{refs}

\end{document}